\documentclass[preprint,showpacs,preprintnumbers,aps,pra,floatfix]{revtex4-1}

\usepackage{bm}
\usepackage{graphicx}
\usepackage{amsmath}
\usepackage{amssymb}
\usepackage[usenames]{color}
\usepackage{subfigure}
\usepackage{amsfonts,amstext,graphics,graphicx}
\usepackage{color}

\graphicspath{{figs/}{figs_lo/}}

\def\sgn{{\;{\rm sgn}}}
\def\<{{\langle}}
\def\>{{\rangle}}
\def\dfr{\displaystyle\frac}
\def\p{\partial}
 
\def\bq{\begin{equation}}
\def\eq{\end{equation}}
\def\bqy{\begin{eqnarray}}
\def\eqy{\end{eqnarray}}
\def\bqyn{\begin{eqnarray*}}
\def\eqyn{\end{eqnarray*}}

\def\calc{{\cal C}}

\def\cale{{\cal E}}

\def\calk{{\cal K}}

\def\calz{{\cal Z}}

\def\cD{{\cal D}}

\def\al{\alpha}
\def\be{\beta}

\def\de{\delta}

\def\ga{\gamma}

\def\si{\sigma}
\def\ze{\zeta}
\def\th{\theta}
\def\om{\omega}

\def\ka{\kappa}
\def\la{\lambda}
\def\La{\Lambda}

 \def\Om{\Omega}
\def\ph{\phi}

\def\varep{\varepsilon}
\def\vep{\varepsilon}

\def\R{\mathbb{R}}

\def\N{\mathbb{N}}
\def\T{\mathbb{T}}
\def\Z{\mathbb{Z}}

\def\A{\mathfrak{A}}
\def\J{\mathfrak{J}}
\def\T{\mathfrak{T}}

\def\M{\mathfrak{M}}
\def\P{\mathfrak{P}}
\def\V{\mathfrak{V}}

\newcommand{\comment}[1]{}
\newcommand{\mathnotation}[2]{\newcommand{#1}{\ensuremath{#2}}}

\mathnotation{\ldef}{\mathrel{\raisebox{.069ex}{:}\!\!=}}
\mathnotation{\rdef}{\mathrel{=\!\!\raisebox{.069ex}{:}}}	 
\mathnotation{\dint}{\,{\mathrm{d}}}		 
\mathnotation{\onehalf}{\tfrac{1}{2}}		 
\mathnotation{\qq}{\theta}

 \def\Xint#1{\mathchoice
{\XXint\displaystyle\textstyle{#1}}%
{\XXint\textstyle\scriptstyle{#1}}%
{\XXint\scriptstyle\scriptscriptstyle{#1}}%
{\XXint\scriptscriptstyle\scriptscriptstyle{#1}}%
\!\!\int}
\def\XXint#1#2#3{{\setbox0=\hbox{$#1{#2#3}{\int}$ }
\vcenter{\hbox{$#2#3$ }}\kern-.5\wd0}}

\def\dashint{\Xint-}

\begin{document}

\title{Continuum Hamiltonian Hopf Bifurcation I}
\author{Philip J. Morrison}
\email{morrison@physics.utexas.edu}
\affiliation{Department of Physics \& Institute for Fusion Studies, University of Texas, Austin 78712-0262, USA}
 \author{George I. Hagstrom}
\email{georgehagstrom@nyu.edu}
\affiliation{CIMS, New York University, New York, NY, USA}
%

\begin{abstract}

Hamiltonian bifurcations in the context of noncanonical Hamiltonian matter models are described.  First, a large class of 1 + 1 Hamiltonian multi-fluid models is considered.  These models have linear dynamics  with discrete spectra, when linearized about homogeneous equilibria, and these spectra have counterparts to the steady state and Hamiltonian Hopf bifurcations when equilibrium parameters are varied.  Examples  of fluid sound waves and  plasma and gravitational streaming  are treated in detail.   Next, using these 1 + 1 examples  as a guide,   a large class of 2 + 1 Hamiltonian systems is introduced, and Hamiltonian bifurcations with continuous spectra are examined.   It is shown how to attach a signature to such continuous spectra,  which facilitates the description of the continuous Hamiltonian Hopf bifurcation.  This chapter  lays the groundwork for Kre\u{i}n-like theorems associated with  the CHH bifurcation  that are more rigorously discussed in our companion chapter \cite{chaptII}. 
\end{abstract}

\maketitle

\section{Introduction}
\label{sec:intro}

\begin{figure}[htb]
  \begin{center}
    \subfigure[ ]{\includegraphics[width=1.9in,angle=0]{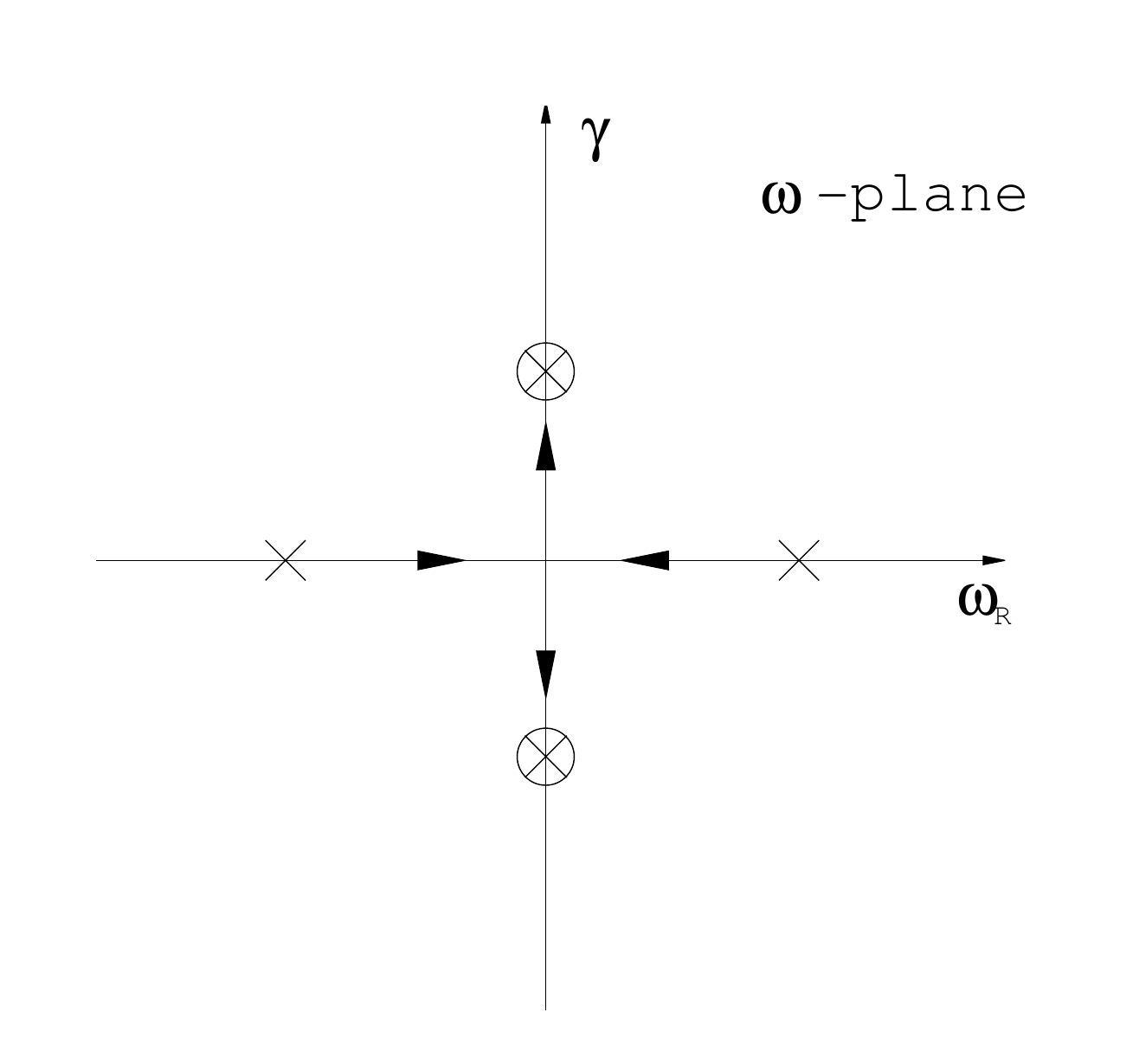}}
    \subfigure[]{\includegraphics[width=2.6in,angle=0]{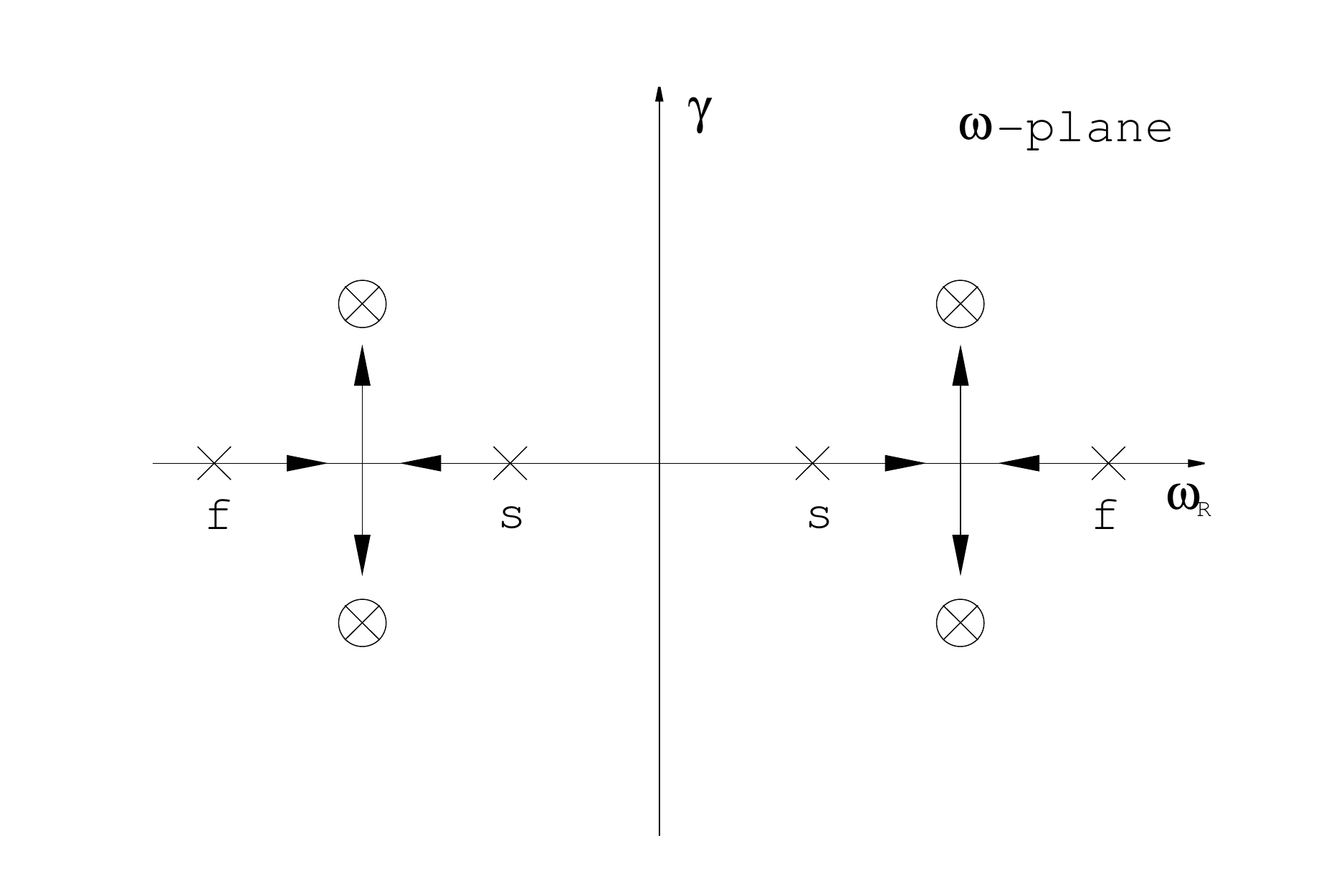}}
  \end{center}
  \caption{Hamiltonian bifurcations with frequency $\om=\om_R+i\ga$. a) Steady state bifurcation with doublet bifurcating through the origin. b) Hamiltonian Hopf bifurcation  showing  Kre\u{i}n collision with  quartet; slow modes (s) have opposite energy signature from fast (f).}
\label{Hbif}
\end{figure}

A common bifurcation to instability, one that occurs in so-called natural Hamiltonian systems that have Hamiltonians composed of the sum of kinetic and potential energy terms,  happens when under a  parameter change the potential energy function changes from positive to negative curvature.   In such a bifurcation, pairs of pure imaginary eigenvalues corresponding to real oscillation frequencies  collide at zero and transition to pure imaginary, corresponding to growth and decay.  This  behavior, which can occur in general Hamiltonian  systems and is termed the steady state (SS) bifurcation, is depicted in the  complex frequency $\om=\om_R + i\ga$ plane in Fig.~\ref{Hbif}a.  Alternatively,   The Hamiltonian Hopf (HH) bifurcation is the generic bifurcation that occurs in Hamiltonian systems when pairs of nonzero eigenvalues collide in the so-called Kre\u{i}n collision \cite{krein2} between eigenmodes of positive and negative signature, as depicted in Fig.~\ref{Hbif}b.  Such bifurcations occur in a variety of mechanical systems \cite{cushman,meer};  however,   HH  bifurcations also occur in infinite-dimensional systems with discrete spectra.  In fact, one of the earliest such bifurcations was identified in the field of plasma physics \cite{sturrock} for streaming instabilities,  where signature was associated with the sign of the dielectric  energy,  and this  idea  made its way into fluid mechanics \cite{cairns,mackay}.  Streaming  instabilities were  interpreted in the noncanonical Hamiltonian context in \cite{morrison90,kueny}, where  signature was related to the sign of the oscillation energy in the stable Hamiltonian normal form \cite{williamson,weierstrass} (see \ref{stabNF} below).

The purpose of this chapter and its companion \cite{chaptII}  is to describe Hamiltonian bifurcations in the noncanonical Hamiltonian formalism (see \cite{morrison98}), which is the natural form for a large class of matter models  including those that describe  fluids and plasmas.  Particular emphasis is on the continuum Hamiltonian Hopf (CHH) bifurcation, which is terminology we introduce for particular bifurcations that arise in Hamiltonian systems when there exists a continuous spectrum.  There also exist a continuum steady state (CSS) bifurcation, but this will only be mentioned in passing.   A difficulty presents itself when attempting to generalize Kre\u{i}n's theorem,  which states that a necessary condition for the bifurcation to instability is that the colliding eigenvalues of the HH bifurcation have opposite signature, to systems with continuous spectra.   This difficult arises because `eigenfunctions' associated with the continuous spectrum are not normalizable, in the usual sense, and consequently obstacles have to be overcome to define signature for the continuous spectrum.  This was done first in the context of the Vlasov equation in \cite{MP92,morrison00a} and for fluid shear flow in \cite{BM02}.    Given this definition of signature, it become possible in \cite{hagstrom,hagstromCL} to define the CHH,  a meaningful generalization of the HH bifurcation.  

In the present chapter  we motivate and explore aspects of the CHH, which are picked up in our companion chapter \cite{chaptII}.  To this end we describe in Secs.~\ref{sec:Discrete} and \ref{sec:theories}  large classes of Hamiltonian systems that possess discrete and continuous spectra when linearized about equilibria.  These classes  are noncanonically Hamiltonian,  as is the case in general  for  matter models in terms Eulerian variables.  For  a general field variable $\Psi$ that represents the state of such a  system,  a noncanonical Hamiltonian dynamical system has the form
\bq
\Psi_t=\{\Psi, H\}= {\J}\,  \frac{\de H}{\de \Psi}\,,
\eq 
where $H[\Psi]$ is the Hamiltonian functional and $\{\, ,\, \}$ the Poisson bracket  defined by 
\bq
\{F,G\}=\int\!d\mu\,  \frac{\de F}{\de \Psi}\,  {\J}[\Psi] \, \frac{\de G}{\de \Psi}\,.
\label{genPB}
\eq
In general one may consider  a $\mu + 1$ multicomponent theory, i.e., $\Psi(\mu,t)=(\Psi^1, \Psi^2, ....)$,  with ${\J}$ being an operator that makes (\ref{genPB}) a Lie algebra realization on functionals (observables).  Because the operator ${\J}$ need not have the canonical form,  may depend on $\Psi$,  and may possess degeneracy this structure was referred to in \cite{MG80} as noncanonical.  Because of the degeneracy, the Poisson bracket of (\ref{genPB}) possesses Casimir invariants $C[\Psi]$ that satisfy
\bq
\{C,F\}\equiv 0 \qquad \forall \ F\,. 
\eq
We refer  the reader to  \cite{morrison98,marsden} for further details. 

In   Sec.~\ref{sec:Discrete} we consider  a class of 1+1 multi-fluid theories, that possess discrete spectra when linearized about homogeneous equilibria.  The linearization procedure along with techniques for canonization and diagonalization, i.e., transformation to conventional canonical form and transformation to the stable normal form, respectively, are developed.  Then,  specific examples are considered that display both SS and HH bifurcations.  In Sec.~\ref{sec:theories} we consider a class of   2+1 theories.  The class is described and the CHH bifurcation for the particular case of the Vlasov-Poisson system is discussed.  Relationship to the results of Sec.~\ref{sec:Discrete}  is  shown by introducing the waterbag model, which is one way of discretizing  the continuous spectrum, and motivates our definition of the CHH bifurcation.  Finally, in Sec.~\ref{sec:conclu1}, we summarize and introduce the material  that will be treated in \cite{chaptII}.

\section{Discrete Hamiltonian bifurcations}
\label{sec:Discrete}

We first describe a class of Hamiltonian theories of fluid type that have equilibria with discrete spectra.   Three examples are considered that demonstrate the occurrence of Hamiltonian bifurcations like those of  finite-dimensional systems.  In the last example of Sec.~\ref{sssec:jeans},   the HH  bifurcation is seen to arise in the context of streaming. 

\subsection{A class of 1 + 1 Hamiltonian multi-fluid theories}
\label{ssec:mulitftheo}

For our purposes here it sufficient to consider a class of 1+1 theories of Hamiltonian  fluid type.  These theories have space-time independent variables $(x,t)$, where $x\in\mathbb{T}\subset\R$, where $\mathbb{T}=[0,2\pi)$, on which we assume spatial periodicity for  dependent variables of fluid type, $\Psi=(\rho_1, \rho_2,  \dots u_1, u_2, \dots)$, where $\rho_{\al}(x,t)$    and $u_{\al}(x,t)$  are the density and velocity fields,  respectively,  with $\al=1,2,\dots, M$.  These fields will be governed  by a coupled set of ideal fluid-like equations generated by a Hamiltonian with a  noncanonical Poisson bracket. 

The noncanonical Poisson bracket for the class is obtained from that for the ideal fluid \cite{MG80, morrison82} reduced to one spatial dimension,
\bq
\{F,G\}= \sum_{\al=1}^M\int_{\mathbb{T}}\!dx
\left( \frac{\de G}{\de \rho_{\al}}
\p  \frac{\de F}{\de u_{\al}} 
-
\frac{\de F}{\de \rho_{\al}}
\p   \frac{\de G}{\de u_{\al}}
\right)\,,
\label{mfPB}
\eq
where the shorthand $\p:= {\p}/{\p x}$ is used and    $\de F/\de u_{\al}$ and $\de F/\de \rho_{\al}$ are the usual functional (variational ) derivatives (see e.g.\ \cite{morrison98}).  We consider Hamiltonian functionals of the following form:
\bq
H[\rho_{\al},u_{\al}]=\sum_{\al=1}^M\int_{\mathbb{T}}\!dx\left(\frac1{2}\rho_{\al} u^2_{\al} + \rho_{\al} U_{\al}(\rho_{\al} ) 
+ \frac1{2} \rho_{\al} \Phi
\right)
\label{mfB}
\eq
where the internal energy per unit mass, $U_\al$, is arbitrary but often taken to be $U_\al=\ka \rho^{(\ga-1)}/(\ga -1)$ where $\ka$ and the polytropic index $\ga$ are positive constants.  The coupling between the fluids is included by means of a  field $\Phi$ that satisfies
\bq
\Phi(x,t)=\sum_{\be=0}^M\P[\rho_{\be}]
\label{mfH}
\eq
where $\P$ is a symmetric  pseudo-differential operator,   $\int_{\mathbb{T}}dx\, f\P[ g]= \int_{\mathbb{T}}dx\,  g\P[f]$, and an arbitrary constant term $\rho_0$ has been included on the right hand side of (\ref{mfH}).

From (\ref{mfH}) we obtain
\bq
 \frac{\de H}{\de \rho_{\al}}=  \frac{u_{\al}^2}{2} + h_{\al} + \Phi
 \quad {\rm and }\quad 
 \frac{\de H}{\de u_{\al}}= \rho_{\al} u_{\al}
 \label{funcDr}
 \eq
where the enthalpy $h_{\al} =  \p( \rho_{\al}U_{\al})/\p \rho_{\al}
$ and the pressure of each fluid is given by $p_{\al}=\rho^2_{\al} \p U_{\al}/\p \rho_{\al}$.  Using (\ref{funcDr}) with (\ref{mfPB}), gives
\bqy
\frac{\p \rho_{\al}}{\p t}&=& \{\rho_\al , H\}= -{\p (\rho_{\al} u_{\al})}\,,
\nonumber\\
\frac{\p u_{\al}}{\p t}&=&  \{u_\al, H\}= - u_{\al} \p u_{\al} - \p p_{\al}/\rho_{\al} -\p \Phi\,,
\nonumber
\eqy
which are a system of fluid equations coupled through $\Phi$ alone. 

The noncanonical bracket of (\ref{mfPB}) is degenerate and possesses the following Casimir invariants
\bq
C^{\rho}_{\al}=\int_{\mathbb{T}}\!dx\, \rho_\al\quad {\rm and}\quad
C^{u}_{\al}=\int_{\mathbb{T}}\!dx\, u_\al\,, \quad  \al=1, \dots, M\,.
\label{mcas}
\eq
These invariants satisfy $\{C^{u,\rho}_{\al}, F\}\equiv 0$ for all functionals $F$.  The physical significance of these Casimirs can  be traced back to  the Liouville theorem of kinetic theory \cite{morrison87} (cf.\ Sec.~\ref{sssec:WB}).

\subsubsection{Equilibrium and stability}
\label{sssec:eqstab}

Because of the existence of the Casimir invariants, the Hamiltonian is not unique and, consequently, equilibria possess a variational principle since
\[
0= {\J}[\Psi] \, \frac{\de H}{\de \Psi}= {\J}[\Psi] \, \frac{\de F}{\de \Psi}\,.
\]
where $F=H + C$.   That is, ${\de F}/{\de \Psi}=0   \Rightarrow   \Psi_t=0$. 
In the present context this amounts to 
$
\de F=\de \left(H +\sum_{\al} \la^{\rho}_{\al} C^{\rho}_{\al} + \la^{u}_{\al} C^{u}_{\al}\right)= 0
$,  with Lagrange multipliers $\la^{\rho,u}_{\al}\in \R$,  or 
\bq
\frac{\de F}{\de \rho_{\al}}=  \frac{u_{\al}^2}{2} + h_{\al} + \Phi +
\la^{\rho}_{\al}=0
 \quad \mathrm{and} \quad
 \frac{\de F}{\de u_{\al}}= \rho_{\al} u_{\al} + \la^{u}_{\al}=0\,.
\label{mfe1}
\eq
Equations (\ref{mfe1})  have the  equilibrium solution $\Phi_e\equiv0$, and  $\rho_{\al}^e\in\R^{>0}$ and $u_{\al}^e\in\R$.  

Expansion around such equilibria give a  linear dynamical system.  Because the equilibria  of interest are homogeneous we can use the following expression en route to linearization:
\[
\rho_{\al}= \rho_{\al}^e + \sum_{k\in\Z}\rho^{\al}_k(t)\,  e^{ikx} 
 \quad \mathrm{and}\quad
 u_{\al}= u_{\al}^e + \sum_{k\in\Z}u^{\al}_k(t)\,  e^{ikx} \,,
\]
where the equilibrium constants $(\rho_{\al}^e ,u_{\al}^e )$ could be absorbed into the sum by redefinition of the $k=0$ terms.   For linearization we expand in the smallness of $(\rho^{\al}_k ,u^{\al}_k )$.

Functionals of $(\rho^{\al}_k ,u^{\al}_k)$ can be mapped onto functions of the Fourier components by insertion of the Fourier series, i.e., 
\[
F[\rho_{\al},u_{\al}] 
=f(\rho_0^{\al},\rho^{\al}_{\pm 1}, \rho^{\al}_{\pm 2}, \dots; u_0^{\al},u^{\al}_{\pm 1}, u^{\al}_{\pm 2},\dots)
\]
and this transformation (for our purposes) can be considered invertible upon using
\[
u_k^{\al}=\frac1{2\pi}\int_{\mathbb{T}}\! dx \, u_{\al}(x) \, e^{-ikx}\,.
\]
Functional derivatives can also be expanded, e.g., 
\bq
\frac{\de F}{\de u_{\al}} = \sum_{k\in\Z} \left(\frac{\de F}{\de u_{\al}}\right)_k \, e^{ikx}
\quad\mathrm{and}\quad
\left(\frac{\de F}{\de u_{\al}}\right)_{-k}\!\!= \frac1{2\pi} \,\frac{\p f}{\p  u^{\al}_k}\,,
\label{fnder}
\eq
where the second equality follows from   the chain rule (see,  e.g., \cite{tassi1}). 
Using (\ref{fnder}) and its counterpart for $\rho_{\al}$, the bracket of (\ref{mfPB}) becomes
\bq
[f,g]=\sum_{k\in\Z}\,\sum_{\al=1}^M \frac{i k}{2\pi} 
\left(
\frac{\p g}{\p  u^{\al}_k}\frac{\p f}{\p  \rho^{\al}_{-k}} -\frac{\p f}{\p  u^{\al}_k}\frac{\p g}{\p  \rho^{\al}_{-k}}
\right)
\label{fpb}
\eq
 Observe that in the Poisson bracket of (\ref{fpb}) the Casimir invariants of  (\ref{mcas}), the $k=0$ components of 
 $(\rho_{\al},u_{\al})$,  have been removed, i.e., the bracket has become nondegenerate  in terms of the ostensible dynamical variables $(\rho^{\al}_k ,u^{\al}_k )$.  Geometrically, the choice of equilibrium selects the symplectic leaf on which the dynamics takes place. 
 
 It remains to determine the Hamiltonian.  This is done by inserting the Fourier expansions of $(\rho_{\al},u_{\al})$  into (\ref{mfB}).  From this one obtains  the full nonlinear dynamics in terms of Fourier amplitudes, but since our interest is in bifurcations of the linear dynamics,  we expand in the smallness of the amplitudes to obtain a quadratic form.   Although this can be done in general terms, we prefer to explore  particular cases of equilibrium and stability  in Sec.~\ref{ssec:mfexamples}.  However, before doing so, we makes some general comments about canonization and diagonalization.

\subsubsection{Canonization and diagonalization}
\label{sssec:candia}

The bracket of (\ref{fpb}) is not yet in canonical form. To canonize  we rewrite the sums as follows:
\bqy
[f,g]&=&\sum_{k\in \N}\,\sum_{\al=1}^M \frac{i k}{2\pi}\left[
\left(
\frac{\p g}{\p  u^{\al}_k}\frac{\p f}{\p  \rho^{\al}_{-k}} -\frac{\p f}{\p  u^{\al}_k}\frac{\p g}{\p  \rho^{\al}_{-k}}
\right)\right.
\nonumber\\
&{\ }& \hspace{2.25 cm}
-\left.\left(
\frac{\p g}{\p  u^{\al}_{-k}}\frac{\p f}{\p  \rho^{\al}_{k}} -\frac{\p f}{\p  u^{\al}_{-k}}\frac{\p g}{\p  \rho^{\al}_{k}}
\right)
\right]
\label{pbraw}\\
&=&\sum_{k\in\N}\,\sum_{m=1}^{2M}\left(
\frac{\p f}{\p  q^{m}_k}\frac{\p g}{\p  p^{m}_{k}} -\frac{\p g}{\p  q^{m}_k}\frac{\p f}{\p  p^{m}_k}
\right)\,,
\label{pbcan}
\eqy
where in the second equality the canonical fields $(q^{m}_k,p^{m}_k)$ are obtained as particular real linear combinations of $(\rho^{\al}_{\pm k},u^{\al}_{\pm k})$.   Thus modes, i.e.,  degrees of freedom,  are indexed by the wavenumber $k\in\N$ and an  index $m$ that takes two values for every value of the species index $\al$.  In terms of a choice of  canonical fields,  the Hamiltonian for the linear dynamics is given by the second variation as  $\de^2 F=:2H_L$ and takes the form
\bq
H_L=\sum_{k,\ell\in \N}\, \sum_{m,n=1}^{2M} z_k^{m}\,   {\A}_{mn}^{k \ell} \,  z_{\ell}^{n}\,,
\label{mfham}
\eq
where $z_k^{m}:=(q^{m}_k,p^{m}_k)$ and the matrix ${\A}^{mn}_{k\ell}$ depends on the 
specific values of the  equilibrium parameters $(\rho_{\al}^e ,u_{\al}^e )$.

Given the system with Hamiltonian in the form of (\ref{mfham}), it remains to effect a canonical transformation to a normal form.  For example, if the system is linearly  stable then there exists a canonical transformation 
$(q^{m}_k,p^{m}_k) \leftrightarrow (Q^{m}_k,P^{m}_k)$, from real variables to real variables, where the Hamiltonian becomes  the following  in terms of the new canonical coordinates
\bq
H_L^s=\frac1{2}\sum_{k\in \N}\, \sum_{m=1}^{2M} \si_k^{m}  \, \omega_k^{m} \, 
\Big((P^{m}_k)^2 + (Q^{m}_k)^2\Big)
\label{stabNF}
\eq
where the frequencies $\omega_k^{m}\in \R^{>0}$ and the signature $\si_k^{m} \in \{\pm 1 \}$.  Thus,  the stable normal form is just an infinite sum over simple harmonic oscillators.  Those for which $\si_k^{m} =- 1$ are negative energy modes, stable oscillations with negative energy (Hamiltonian).  It is important to emphasize that even though the energy is negative, the system is stable.  For finite-dimensional systems,  the method for constructing the canonical transformation to normal coordinates $(Q^{m}_k,P^{m}_k)$ is treated in standard texts and this method carries over.  However, when negative energy modes exist the method is complicated somewhat and, although well-known in Hamiltonian dynamics lore, is not usually treated in physics texts.  An accessible treatment is given in \cite{tassi1},  where it is applied in a plasma physics context.

\subsection{Examples}
\label{ssec:mfexamples}

In order to make the ideas discussed in Secs.~\ref{sssec:eqstab} and \ref{sssec:candia} more concrete we consider a few examples that demonstrate explicitly canonization,  diagonalization, and Hamiltonian bifurcations to instability in the context of multi-fluid models; in particular,  the HH bifurcation will emerge for particular modes indexed by $(k,\al)$,  just as it appears in finite-dimensional systems. 

\subsubsection{Sound waves and multiplicity}
\label{sssec:sound}

Consider first the case of a single fluid with an equilibrium state given by $\rho_e$ some positive constant and 
$u_e\equiv 0$.  The linear Hamiltonian is evidently
\bq
H_L= \frac1{2}\int_{\mathbb{T}}\! dx \, 
\left(
\rho_e (\de u)^2 + c_s^2 (\de \rho)^2 /{\rho_e}
\right)
=
\pi\sum_{k\in\Z}\, 
\left(
\rho_e |u_k|^2 + {c_s^2}\, (|\rho_k|^2 /{\rho_e}
\right)\,, 
\nonumber
\eq
where $c_s^2=p_{\rho_e}=\rho_e(\rho U)_{\rho \rho} (\rho_e)$ is the sound speed.   The appropriate Poisson bracket is 
is that of (\ref{pbraw}) with a single $\al$-term.  

With some thought canonization and diagonalization is possible in a single step, but we will proceed in a direct manner by assuming the canonical coordinates are  
\[
q^1_k=\sqrt{2\pi}(u_k + u_{-k}) \quad \mathrm{and}\quad q^2_k=\sqrt{2\pi}(\rho_k + \rho_{-k})
\]
with corresponding momenta
\[
p^1_k=\frac{\sqrt{\pi/2}}{ik} (\rho_k - \rho_{-k}) 
\quad \mathrm{and}\quad p^2_k= \frac{\sqrt{\pi/2}}{ik}(u_k - u_{-k})\,.
\]
A simple chain rule calculation takes (\ref{pbraw}) into the following:
\bq
[f,g]=\sum_{k\in\N} \, \sum_{m=1}^2  \left(
\frac{\p f}{\p  q^{m}_k}\frac{\p g}{\p  p^{m}_{k}} -\frac{\p g}{\p  q^{m}_k}\frac{\p f}{\p  p^{m}_k}
\right)\,.
\label{cansnd}
\eq
Observe in (\ref{cansnd}) that indexing a degree of freedom by $k\in \N$ requires a multiplicity index $m$.  Each mode, which is described by an amplitude and a phase,  constitutes a single degree of freedom; a single degree of freedom is thus two dimensional and,  consequently each mode corresponds to two eigenvalues.  For a stable degree of freedom these eigenvalues correspond to two frequencies, one the negative of the other.   Here we have multiplicity, the  reason for which will become mentioned when we diagonalize.  

Now,  using
\[
u_k=\frac{1}{2\sqrt{2\pi}}\left( q_k^1 + 2ik \,p^2_k\right)
\quad \mathrm{and}\quad
\rho_k=\frac{1}{2\sqrt{2\pi}}\left( q_k^2 + 2ik \,p^1_k\right)\,,
\]
valid with $k\rightarrow-k$, in the Hamiltonian $H_L$  gives
\[
H_L[q,p]= \frac1{4}\sum_{k\in\N}\Big(
\rho_e|q_k^1|^2 + 4k^2 c_s^2\, |p_k^1|^2/{\rho_e}
+ c_s^2 |q_k^2|^2/{\rho_e} +4k^2\rho_e\, |p_k^2|^2
\Big)\,.
\]
The normal form is achieved upon substitution of the following canonical transformation:
\[
q^1_k=\sqrt{\frac{2kc_s}{\rho_e}}\,Q^1_k\,, \quad
p^1_k=\sqrt{\frac{\rho_e}{2kc_s}}\,P^1_k\,, \quad
q^2_k=\sqrt{\frac{2k\rho_e}{c_s}}\,Q^2_k\,, \quad
p^2_k=\sqrt{\frac{c_s}{2k\rho_e}}\,P^2_k\,, 
\]
i.e., $H_L$  becomes
\bq
H_L[Q,P]=   \frac1{2}\sum_{k\in\N}\, \sum_{m=1}^2 kc_s\Big((Q^m_k)^2 + (P^m_k)^2\Big)\,.
\label{soundnorm}
\eq
This is the sought after normal form where the frequency of all modes is $kc_s$ as appropriate for sound waves. 

We close this example with a few comments.  First, for a given wavelength as determined by $k\in\N$ there are in fact two modes, one that propagates to the right and one to the left.  This is accounted for by the multiplicity index, $m$. 
In obtaining this normal form we have assumed $c_s^2=p_{\rho}>0$, which can be traced back to a property of $U(\rho)$ and is in essence  Le Ch\^atelier's principle of thermodynamics, viz., that pressure increases upon compression. If we had some exotic fluid for which this was not the case, then the the system would be unstable and the normal form of (\ref{soundnorm}) would not be achievable.  Imagine that the equilibrium parameter $\rho_e$ can be varied and that at some critical value $c_s^2$ makes a transition from positive to negative.  Since a mode frequency  $\om=kc_s$ it is evident that this transition happens at zero frequency and, consequently, is a  SS bifurcation (See Fig.~\ref{Hbif}a).  Moreover, because of the multiplicity this is a degenerate bifurcation where, for each fixed $k$,  four pure imaginary eigenvalue collide at zero frequency and then transition to four pure real eigenvalues of growing and decaying pairs.  The situation is completely degenerate since this happens for all $k$ values simultaneously.  In Sec.~\ref{ssec:comp} we will see that the HH bifurcation, as depicted in  Fig.~\ref{Hbif}b,  can be transformed to a similar collision with four eigenvalues at zero frequency, but it differs in that after bifurcation one obtains the Hamiltonian quartet, four eigenvalues with both real and imaginary parts, a situation that is sometimes called over stability.

\subsubsection{Counterstreaming ion beams with isothermal electrons}
\label{sssec:counterstreaming}

Next we consider a simple one-dimensional multi-fluid plasma configuration
consisting of two cold counterstreaming ion beams in a neutralizing
isothermal electron background.  A detailed linear, nonlinear, and numerical 
analysis of this problem, from a Hamiltonian perspective, can be found in \cite{kueny,kueny2} and we refer the reader  there for further details.

The dynamical system of interest  in dimensionless form is given by 
\bqy
&&\dfr{\p u_\al}{\p t}+u_\al{\p u_\al} 
+ {\p\ph} =0, 
\nonumber\\
&&\dfr{\p \rho_\al}{\p t}+ {\p} 
\left(\rho_\al u_\al\right)=0,
\nonumber\\
&& {\p^2\ph} =e^\ph - \rho_+ - \rho_- \,,
\label{ions:poi}
\eqy
where $\al\in\{\pm\}$ labels each ion stream with velocity $u_{\al}$.   Here $\rho_{\al}$ represents a dimensionless number density instead of mass density.    Equation (\ref{ions:poi}), Poisson's equation,  is a constraint and  in principle the electrostatic potential  can in principle solved as
$\ph\left(\rho_+,\rho_-\right)$,  so that the entire system is
described in terms of the dynamical variables $\rho_\pm$ and $u_\pm$.  As usual the electric field is given by $E=-\p \phi$.  Thus,  this system is of the class described in Sec.~\ref{ssec:mulitftheo}, with (\ref{ions:poi}) a specific case of (\ref{mfH}).   It has the Hamiltonian functional
\[
H=\int_{\mathbb{T}}\! dx \, \left(\sum_{\al}\frac1{2}\rho_{\al} u^2_{\al}
+\int_0^\ph \!\!d\ph\, \ph \,e^\ph+\dfr{1}{2}\left( {\p\ph}\right)^2\right)
\,, 
\]
and Poisson bracket of (\ref{mfPB}).

Homogeneous equilibria follow from $\de F=0$  for $\la^{\rho}_\pm=-{u_e^2}/{2}$ and 
$\la^u_\pm=\mp{u_e}/{2}$, 
which are consistent with an equilibrium of ion streams
of equal density and speed, 
\[
\rho^+_e=\rho_e^-=\dfr{1}{2},\quad\quad u_e^+=-u_e^-=u_e,\quad\quad E_e=\ph_e=0\,, 
\]
that we assume for simplicity.  Thus,  we have a one parameter family of equilibria controlled by $u_e$. 

Linearizing about  this equilibrium state gives the following Hamiltonian for the linear dynamics: 
\bqy
H_L&=&\frac1{2}\int_{\mathbb{T}}\! dx \,  
\Bigg(
\dfr{1}{2}(\de u_+)^2 + \dfr{1}{2}(\de u_-)^2 
+2\,u_e\,\de \rho_+\de u_+ 
\nonumber\\
&&\hspace{3 cm} -2\,u_e\,\de \rho_-\de u_-+(\p \de\ph)^2+(\de\ph)^2
\Bigg)\,. 
\nonumber
\eqy
Observe, the sign of $H_L$ may be either positive or negative, depending
on the perturbation; thus we may have instability, or negative energy modes in the system.
 
Expansion in a Fourier series as in Sec.~\ref{sssec:eqstab}, including the expansion  
$\de \phi=\sum_{k\in\Z} \phi_k e^{ikx}$, and using  the  linearized Poisson equation  (\ref{ions:poi})  gives
$\phi_k={N_k }/({1+k ^2})$,  where $N_k := \rho_k^+ + \rho_k^{-}$.  With this expression,  
the energy $H_L$ becomes
\bqy
H_L&=&\dfr{\pi}{2}\sum_{k\in\N} \bigg(
  |u_k^+|^2+|u_k^-|^2 
  \label{del2f2}\\
&& \hspace{2 cm}  +2u_e\big(\rho_k^+u_{-k}^+-\rho_k^-u_{-k}^-+\mbox{c.c.}\big)
+2\dfr{\left|N_k \right|^2}{1+k ^2}
\bigg)\,, 
\nonumber
\eqy
where c.c.~denotes complex conjugate.  Under  the transformation
\bqy
\rho_k^+&=& \sqrt{\frac{u_e}{\pi}} \frac{k}{2}
\left(p_2- i  q_1\right),\qquad u_k^+=\dfr{1}{2 \sqrt{\pi u_e}} \left(p_1- i q_2\right)\,
\nonumber\\
\rho_k^-&=& \sqrt{\frac{u_e}{\pi}} \frac{k}{2} \left(p_4-i q_3\right),\qquad   
u_k^-=\dfr{1}{2 \sqrt{\pi u_e}}\left(p_3-i q_4\right),  
\label{nvtoqp}
\eqy
with $\rho_{-k}^{\pm}=( \rho_k^{\pm})^*$ and $u_{-k}^{\pm}=(u_k^{\pm})^*$,  the  Poisson bracket  becomes that of (\ref{pbcan}) with $M=2$ 
and the linear Hamiltonian (\ref{del2f2})  becomes 
\[
H_L=\dfr{1}{2}\sum_{k\in\N}\, \sum_{m,n=1}^4\big(
p_k^m\, {\M}^k_{mn}\, p_k^n 
 + 
q_k^m\, {\V}^k_{mn}\, q_k^n \big)\,, 
\]
where
\bqy
{\M}^k&=&\left[
\begin{array}{cccc}
\dfr{1}{2 u_e } &\quad k u_e  &\quad 0 &\quad 0\\[20pt]
ku_e  &\quad \dfr{k ^2 u_e}{1+k ^2} &\quad 0 &\quad
\dfr{k ^2 u_e }{1+k ^2}\\[20pt]
0 &\quad 0 &\quad \dfr{1}{2 u_e } &\quad -k u_e \\[20pt]
0 &\quad \dfr{k ^2 u_e }{1+k ^2} &\quad -k u_e  &\quad
\dfr{k ^2 u_e }{1+k ^2}
\end{array}
\right]\,.
\nonumber\\
{\V}^k&=&\left[
\begin{array}{cccc}
\dfr{k ^2 u_e }{1+k ^2} &\quad k u_e &\quad \dfr{k ^2 u_e }{1+k ^2} &\quad
0\\[20pt]
k u_e  &\quad \dfr{1}{2 u_e } &\quad 0 &\quad 0\\[20pt]
\dfr{k ^2 u_e }{1+k ^2} &\quad 0 &\quad \dfr{k ^2 u_e }{1+k ^2} &\quad
-k u_e \\[20pt]
0 &\quad 0 &\quad -k u_e  &\quad \dfr{1}{2 u_e }
\end{array}
\right]
\nonumber
\eqy
Thus, in terms of the canonical coordinates of (\ref{nvtoqp}) the system is diagonal in $k$, but it remains to transform  the  $4\times4$ block structure, the part corresponding to the multiplicity,  to normal form. 

For values of $u_e$ for which the system is stable, the  diagonalizing  canonical transformation is given explicitly in an Appendix of \cite{kueny}.  The reader is directed there to see how to obtain
\bqy
H_L^s&=&\frac1{2}\sum_{k\in\N}  \Big(
 \om_k^+\big({(P_k^1)^2+(Q_k^1)^2}\big)
-
\om_k^-({\big(P_k^2)^2+(Q_k^2)^2}\big)
\label{del2f4}\\
&&\hspace{1.5 cm} 
+
\om_k^+\big({(P_k^3)^2+(Q_k^3)^2}\big)
-
\om_k^-\big({(P_k^4)^2+ (Q_k^4)^2}\big)
\Big)\,.
\nonumber
\eqy
Evidently for each value of $k$  there exist four modes, two positive energy modes and two negative energy modes.  The symmetry of the equilibrium  facilitates the  calculation of the frequencies, which are given by
\bq
\om_k^{\pm}:=k \left[\dfr{1}{2\left(1+k ^2\right)}+u_e^2
\pm\sqrt{\dfr{1}{4\left(1+k ^2\right)^2}+
\dfr{2u_e^2}{\left(1+k ^2\right)}}\,\right]^{\onehalf}>0\,,
\label{freqs}
\eq
which can be obtained from the plasma fluid dielectric (dispersion) function
\bq
\varepsilon(k,\om)= 1+ \frac1{k^2} - \frac1{2}\left(
\frac1{(\om-k u_e)^2} + \frac1{(\om+k u_e)^2} 
\right)=0\,.
\label{dielectric}
\eq
 From (\ref{freqs}) it is evident that all bifurcations to instability occur through zero frequency as depicted in Fig.~\ref{Hbif}a and in fact are degenerate; i.e., if we fix $k$ and vary $u_e$, then there is a value of $u_e$ at  which $\om_-$, the slow mode,  vanishes and then becomes unstable with pure imaginary eigenvalues, two representing growth and two decay.    Thus, this is another example of a SS  bifurcation that is forced to be degenerate because of the imposed symmetry.    In the next section we will break this symmetry and obtain the HH bifurcation, but for variety we do so in a physically different, yet mathematically similar,  context.

\subsubsection{Jeans instability with streaming}
\label{sssec:jeans}

The widely studied Jeans instability occurs in Newtonian gravitational matter models.  For the present example, we suppose matter is governed by our 1+1 fluid model with two interpenetrating streams.  We refer the reader to \cite{casti} for background material and further details.  The model is the same as that of Sec.~\ref{sssec:counterstreaming} except Poisson's equation is replaced by  
\bq
\p^2 \phi = \rho_++\rho_--\rho_{_\Lambda}\,,
\label{eq:3Dgrav}
\eq
where we incorporate  Einstein's device of introducing
a cosmological repulsion term, which in the Newtonian setting  amounts to introducing 
a   negative  constant gravitational mass of density
$\rho_{_\Lambda}$. The sign change in (\ref{eq:3Dgrav}) accounts for gravitational attraction. 

The equilibrium for this case is similar to  that of Sec.~\ref{sssec:counterstreaming}, except we allow for asymmetry and,  like the equilibrium of Sec.~\ref{sssec:sound}, we allow for pressure in each stream. Specifically, we have the equilibrium constant densities  $\rho^+_e$ and  $\rho^-_e$, such that $\rho^+_e+\rho^-_e=\rho_{\La}$ and $\phi_e=0$, the two stream velocities $u_e^+>0$ and $-u_e^->0$, chosen in opposite directions, and two sound speeds $c_s^{\pm}$.  Upon scaling, these can be reduced to four independent equilibrium parameters:  $u_e^+$,  $u_e^-$, $\beta:=\rho^-_e/\rho^+_e$, and 
$c:=c_s^{-}/c_s^{+}$.

From the results of Secs.~\ref{sssec:counterstreaming} and \ref{sssec:sound} we can immediately write down the linearized Hamiltonian
\bqy
H_L&=&\frac1{2}\int_{\mathbb{T}}\! dx \,  
\Bigg(
{\rho^+_e}(\de u_+)^2 +  {\rho^-_e}(\de u_-)^2 
+2\,u^+_e\,\de \rho_+\de u_+ - 2\,u^-_e\,\de \rho_-\de u_-
\nonumber\\ &&\hspace{1 cm} 
+
  (c_s^+)^2 (\de \rho_+)^2 /{\rho^+_e}
 + (c_s^-)^2 (\de \rho_-)^2 /{\rho^-_e}
 - (\p\de\ph)^2 
\Bigg)\,. 
\nonumber
\eqy
Fourier expansion and  canonization proceeds in the same manner as in the previous examples.  In the case where the equilibrium parameters indicate stability, then diagonalization can be  shown to give a Hamiltonian of the form of (\ref{stabNF}) with $M=2$. 

The frequencies are roots of the following `diagravic' function
\bq
\Gamma(k,\om) = 1 +{\frac{1}{2\left[(\omega-ku_e^+)^2-k^2\right]}}+
{\frac{\beta}{2\left[(\omega+cku_e^-)^2-c^2 k^2\right]}}=0\,
\label{diagravic}
\eq
with two fast modes being positive energy modes and two slow modes being negative energy modes  
(cf.\ Figs.~1 and 3 of \cite{casti}).  In general all four modes are  distinct, but if we symmetrize parameters as in 
Sec.~\ref{sssec:counterstreaming}, then the quartic obtained from  (\ref{diagravic}) becomes biquadratic and is  easily solved,  indicating  degenerate modes of each sign as before.  Evidently, this system  possesses a rich parameter space, and various bifurcations to instability for various $k$-values are possible. In addition to the four parameters above, one can use also use $k$ as a control parameter: we have scaled the system size to $2\pi$, but upon reinstatement this translates into varying $k$.  The Jeans instability is a long wavelength instability, and one can observe the transition to instability as $k$ decreases.  This is immediate if  $c=\be=1$ and $u_e=0$, in which case (\ref{diagravic}) implies $\om^2=k^2-1$.  Using $k$ as the control parameter,   as  the wavelength is increased one sees the instability set in as a degenerate SS bifurcation.    The situation is complicated with the presence of two streams, the subject of this subsection, and the HH bifurcation as depicted in Fig.~\ref{Hbif}b is clearly present  (cf.\ Fig.~2 of \cite{casti}).  This is quite generally the case for fluid systems with steaming equilibria.  In Sec.~\ref{sssec:WB}, we will see how multi-fluid streaming relates to the waterbag distribution of kinetic theory, and we will discuss explicitly the HH bifurcation in this context. 
  
\subsection{Comparison and commentary}
\label{ssec:comp}

It is evident from the discussion of Sec.~\ref{ssec:mfexamples} that a requisite for determining a HH bifurcation is the identification of the energy for the linear system.  In the context of noncanonical Hamiltonians systems this  naturally comes from  second variation $\de^2F$,   the Hamiltonian for the linear system.  Sometimes `energy' expressions are obtained by direct manipulation of the linear equations of motion, as done,  for example,  in the original MHD energy principle paper \cite{bernstein58b}, but this procedure can obscure  the notion of signature.  For example, a system of two simple harmonic oscillators conserves  $\om_1(q_1^2 + p_1^2)\pm \om_2 (q_2^2 + p_2^2)$
for both signs and either might be obtained by manipulation of the equations of motion.  The unambiguous sign for the correct energy is uniquely given by $\de^2F$;  this is important because this sign can drastically affect the behavior of the system when dissipation or nonlinearity is considered.  For example, a system with a negative energy mode can become unstable to arbitrarily small deviations from the equilibrium when nonlinearity is added.  (See Cherry's example as described in \cite{morrison90,mp89}.)

In the plasma literature, other definitions of energy are usually considered, e.g, in the context of streaming instabilities  the dielectric energy, which is proportional to $\om |E|^2 \p \varep/\p \om $, where $E$ is the electric field amplitude,  is incorporated.  This expression was originally derived  by von Laue \cite{laue} for the energy content in a dielectic medium by tracking the energy input due to an external agent.  However, we have seen how it arises  from $\de^2F$, and only then can one be assured that it represents a quantity conserved by the linear dynamics.  In fact, for our general multi-fluid model the  $\varep(k,\om)$ takes the form $\varep(k,\om)= 1 +\sum_{\al}\chi_{\al}(k,\om)$, with a contribution $\chi_{\al}$ from each fluid, e.g., that for counterstreaming  and Jeans are (\ref{dielectric}) and (\ref{diagravic}), respectively,   (also cf.\ (\ref{wbvareps}) below) and it can be seen in general that $\de^2 F\sim\om |E|^2 \p \varep/\p \om$.
For neutral modes embedded in the continuous spectrum of Vlasov theory (cf.\ \cite{chaptII}) the formula $\om |E|^2 \p \varep/\p \om$  remains valid \cite{ms94,sm94},   but this formula is incorrect for excitation of the continuous spectrum as shown in  \cite{MP92}, where the correct alternative formula was first derived, and the notion of signature for the continuous spectrum was defined. 

Sometimes energy is  defined in terms of the Lagrangian displacement variable, as was done by manipulation of the  linear equations of motion in \cite{bernstein58b,frieman60}.   Such expressions can also be obtained by expansion of an appropriate Hamiltonian, $\de^2H$.  It was shown in \cite{morrison90,morrison98} that this procedure gives an expression that is essentially equivalent to $\de^2F$. See \cite{amp2a} for a recent general discussion in the context of MHD. 

In conventional Kre\u{i}n theory, the signature is defined in terms of the Lagrange bracket (see, e.g., \cite{moser58}).  However it is a simple matter to see that this corresponds to the normal form definition \cite{mackay2,morrison90}, which follows by comparison of terms in the diagonalization procedure  (cf.\ \cite{whittaker,tassi1}).  In \cite{morrison90} it was argued that all these definitions of signature, that  using, the dielectric energy, $\de^2F$, $\de^2H$, and the Lagrange bracket,  are essentially the same when they are meaningful. 

One ostensible difference between the HH and SS  bifurcations is that the latter occurs at zero frequency.  
However, one can effect a time-dependent canonical transformation so that all four HH  eigenvalues of Fig.~\ref{Hbif}b  collide at zero frequency.   To see this consider one of the  stable degrees of freedom, which  has a  contribution to the Hamiltonian in action-angle variables $(\theta_f, J_f)$ given by  $H_f=\om_f J_f$, where $\omega_f$ depends on the bifurcation control parameter and takes the value $\om^*$ at the bifurcation point.  Using the mixed variable generating function $F_2$ to transform to new canonical variables $(\Theta,\mathcal{J})$, 
\bq
F_2(\theta_f,\mathcal{J}_f)= (\theta_f- \om^* t)\mathcal{J}\,, 
\label{shift}
\eq
we obtain
\[
\Theta=\frac{\p F_2}{\p \mathcal{J}}= \theta_f- \om^* t\quad \mathrm{and}\quad 
J=\frac{\p F_2}{\p \theta_f}=\mathcal{J}\,, 
\]
which amounts to moving into a rotating  coordinate system with new Hamiltonian 
\[
\bar H= H +\frac{\p F_2}{\p t}=(\om_f   - \om^*)  \mathcal{J}\,. 
\]
Thus, in the new frame, the frequency $\om_f   - \om^*$ vanishes at the bifurcation point.  At bifurcation the companion mode $\om_s$ has the same value $\om^*$, consequently,  a similar transformation will bring this mode to zero frequency at bifurcation.  At first glance one might think this has made the HH bifurcation identical to a degenerate SS bifurcation, but the behavior of the two beyond the bifurcation point is different.  The degenerate SS bifurcation transitions to two purely growing and two purely decaying eigenvalues, while the HH transforms to over stability, i.e., it  obtains  a quartet structure immediately upon bifurcation.   (Note, one could argue that the frame shift could be a function of control parameter, but with this line of reasoning all bifurcations could be make to look like SS bifurcations, even into the nonlinear regime.)  
The frame shift of (\ref{shift}) is identical to a Galilean shift that can be done for fluid and plasma theories in order to bring modes to zero frequency at bifurcation.  This artifice is  used in the development of the single-wave model \cite{SW}  and will be considered in \cite{chaptII}. 

The connection between degeneracy and symmetry is well-known and there is a very large literature on  bifurcations in finite-dimensional Hamiltonian systems with  symmetry (see, e.g,  \cite{dellnitz} for an entryway).  In our examples above we have seen, as one might expect, that this is also the case for infinite systems with discrete  spectra.  In fact, it is quite common for the dispersion relation to factor as a consequence of symmetry \cite{tassi2}.  However, systems with symmetry and continuous spectra are less well-studied, but counterparts exist, e.g., the degeneracy of the SS bifurcation of Jeans inability with $u_e=0$ of Sec.~\ref{sssec:jeans} has a CSS counterpart when described by the Vlasov system (see Sec.~IIID of \cite{SW}).  

\section{Continuum Hamiltonian  bifurcations}
\label{sec:theories}

Now we turn to the general class of 2+1 Hamiltonian mean-field theories in which the linear theory around equilibria possess a continuous spectrum.   This is followed by the exposition of the two-stream instability in the Vlasov-Poisson equation, which is a standard example of the CHH bifurcation. Next we introduce the waterbag reduction of the
Vlasov-Poisson equation and use it to connect the two-stream instability to Kre\u{i}n-bifurcations in the corresponding waterbag model, linking this section to Sec.~\ref{sec:Discrete}.

\subsection{A class of 2 + 1 Hamiltonian mean field theories}
\label{ssec:badass}

We begin with the class of 2+1 Hamiltonian field theories introduced in  \cite{morrison03}, which have  with a single dynamical variable,  $f(q,p,t)$,    a time-dependent density on the phase space  variables $z:=(q,p)$.    The density satisfies a transport  equation, 
\bq
\frac{\partial f}{\partial t} + [f ,\cale]=0\,,
\label{eq:den.eq}
\eq 
where the bracket
$[f,g]= f_{q} g_{p} - g_{q} f_{p}$   is the Poisson bracket for a single particle with particle  energy $\cale$,
which will depend globally on $f$. Equation (\ref{eq:den.eq}) is therefore a mean field theory, where $f$ is a density of particles in phase space that generates $\cale$ and is advected along the single particle trajectories that result from 
$\cale$.  The resulting equations are typically quasilinear partial integrodifferential equations. 
We assume that the particle energy arises from a  Hamiltonian functional of the form $H[f]= H_1 + H_2 + H_3 +\dots$, where generally $H_{n}$ is the $n$-point energy, e.g., 
\bq
H_1[\ze] = \int_{\calz}\!\! d^2z\,  h_1(z) \, f(z) \,,\quad
H_2[\ze] =\frac1{2} \int_{\calz} \!\! d^2z \int_{\calz} \!\! d^2z' 
f(z)\, h_2(z,z') \, f(z')\,,
\nonumber
\eq
with $h_{1}$ and $h_{2}$ being  interaction kernels. Here we will only consider Hamiltonian systems with up to  binary interactions, and we will assume that that $h_2$ possesses the symmetry $h_2(z,z') =
h_2(z',z)$. If $\cale$ is obtained from the field energy by functional
differentiation:
\[
\cale :=\frac{\delta H}{\delta f} = h_1 +  \int_{\calz}\!\! d^{2}z'\,  h_2(z,z') \,
f(z')\,,
\]
then $H[f]$ is a constant of motion for (\ref{eq:den.eq}).

Equation (\ref{eq:den.eq}) with $\cale = \de H/\de f$ is a Hamiltonian field theory \cite{morrison03} in terms of the noncanonical Lie-Poisson bracket  of \cite{morrison80,morrison82}
\bq 
\{F,G\}=\int_{\calz}\!\!d^2z\,  f 
\left[\frac{\de F}{\de f},\frac{\de G}{\de f}\right]\,.
\label{eq:pb}
\eq
This bracket depends explicitly upon $f$, unlike
usual Poisson brackets that only depend on (functional) derivatives of the
canonical variables.  The bracket of
(\ref{eq:pb}) is antisymmetric and satisfies the Jacobi identity, though it is degenerate, unlike canonical brackets.  The equations of motion may be written:
\bq
\frac{\partial f}{\partial t}=\{f, H\}
= - \left[f, \frac{\de H}{\de f}\right]
= - [f, \cale]\,,
\label{eq:eom0}
\eq
where $H=H_{1}+H_{2}+\dots$. 

As mentioned in Sec.~\ref{sec:intro},  degeneracy of the Poisson bracket gives rise to Casimir invariants, 
quantities that are conserved for {\em any} Hamiltonian.  For the bracket of (\ref{eq:pb}) the Casimir invariants are 
$C[f]= \int_{\calz} d^2z \,  \calc(f)$,
where $\calc(\ze)$ is an arbitray function.  The existence of Casimir invariants leads to a foliation of phase space (in this case a function space) with symplectic leaves, which are the level sets of the Casimir invariants and which inherit a symplectic structure from the Lie-Poisson bracket. The evolution of $f$ is restricted to one of these symplectic leaves, and the equations on a single leaf are canonical.  

In addition to the Casimir invariants and the total energy, there may be conserved momenta $P[f]$
generally arising from translation symmetries of the interaction kernels  $h_{1}, h_{2}, \dots$. The system conserves momentum if there exists a canonical transformation of the phase space $\calz$,  
${z=(q,p) \longleftrightarrow \bar{z}:=(\qq,I)}$, 
such that in the new particle coordinates $\bar{z}:=(\qq,I)$, the
interactions $h_{1}$,  $h_{2}$, etc.\  have upon composition with $z(\bar z)$ one of the following two forms:
 \bq 
 h_1\circ z= \bar h_1(I)\,,\quad\quad h_2\circ (z, z') =\bar h_2(I,I',|\qq -\qq'|) 
 \label{eq:vpform} 
\eq 
or
 \bq 
 h_1\circ z= 0\,,\quad\quad h_2\circ (z,z') =\bar h_2(|I-I'|,|\qq-\qq'|) \,.
 \label{euform}
\eq  
  
In the first case  
\[
P[f]= \int_{\calz}\!\!d^2z  \, I\, f(z)\,.
\]
is conserved, while in the second case we have two kinds of translation invariance and thus two components of the momentum
\[
P_1[f]= \int_{\calz}\!\!d^2z  \, I\, f(z) \, \quad {\rm and} \quad P_2[f]= \int_{\calz}\!\!d^2z \,\, \qq\, f(z)\,.
\]
These momenta can be very useful (cf., e.g.,  \cite{BM02}), but they will not be discussed further here.

Equilibrium states have  $f$  a function of the single particle constants of motion only, i.e.,  the single particle energy $\cale$ and possibly momenta.  The example treated here has an  equilibrium that only depends  
on $I$, where $(\qq,I)$ are the action-angle variables corresponding to a given $\cale$.   For this reason we set $f(\qq,I,t) = f_0(I) + \ze(\qq,I,t)$  and then when a choice of $f_0$ is made,  $\ze(\qq,I,t)$ represents  the 
main dynamical variable. The phase space is $\mathcal{Z}= \cD\times \mathbb{T}$, i.e., periodic in $\qq\in  [0,2\pi)=\mathbb{T}$, and $I\in\cD$ where here $\cD=\R$.  Upon substitution of $f = f_0 + \ze$ into $\cale$, both 
of the forms of (\ref{eq:vpform}) and (\ref{euform}) can be written as follows:
\[
\cale[f_0+\ze]=\cale[f_0] + \cale[\ze]=:h(I) + \Phi(\qq,I)\,, 
\]
with
\[
\Phi(\qq,I) = \calk \ze  
\ldef \int_\cD\!\! \dint I'\int_{\mathbb{T}}\!\!\! \dint  \qq'  \, K(I,I',|\qq-\qq'|)
         \,\ze(\qq',I',t)\,,
\]
where $h$ and $K$ are determined by $h_1$ and $h_2$. 
Thus the governing equation is 
\bq
  \ze_t + [f_0,\Phi] + [\ze,h+\Phi]=0\,, 
\label{eq:eom}
\eq
where $[f,g]=f_{\th} g_{I} - g_{\th} f_{I}$ and $\Om(I)=h'$.      Equation (\ref{eq:eom}) will serve as a starting point for our subsequent linear analyses.

\subsection{Example of the continuum Hamiltonian Hopf bifurcation}

All of the models described in Sec.~\ref{ssec:badass} possess  CHH bifurcations; however, here we will concentrate on the Vlasov-Poisson system.   First we describe it, then we make  connection to the multi-fluid results of Sec.~\ref{ssec:mulitftheo} and in this way relate the CHH to the ordinary HH bifurcation. 
 
\subsubsection{Vlasov-Poisson system}
\label{ssec:VPS}

The Vlasov-Poisson equation arises out of (\ref{eq:den.eq}) through definition of the single particle energy $\cale$ and potential $\phi$, where $\cale=p^2/2 +\phi $ and 
\bq
f_{t}=-[f,\cale]=-pf_q+\phi_q f_p 
\quad \mathrm{and}\quad 
 \phi_{qq}=1-\int_{\R}\!f dv 
\,.
\eq
The interaction kernels for this model are:
$h_1= {p^2}/{2}$ and $h_2=|q-q'|$. 

The function $f$ represents the density of a positive charge species in phase space, under the assumption that there is neutralizing background with uniform negative charge density. The particles interact with each-other through electrostatic forces, which are included by the Poisson equation. Under the
identification $q=\theta$, $p=I$, we recover (\ref{eq:vpform}). Arbitrary functions of $p$ alone, $f(q,p)=f_0(p)\equiv f_0(I)$, form an important class of solutions to this model, the spatially homogeneous equilibria. The analog of (\ref{eq:eom}) is:
\[
  \ze_t + p \ze_q - f_0' \phi_q-\ze_p\phi_q=0
  \quad \mathrm{and}\quad
\phi_{qq}=-\int_{\R}\!dp \, \ze \,.
\]

Upon linearizing the Vlasov-Poisson system around a homogeneous stable equilibrium, i.e., dropping the nonlinear term  $\ze_p\phi_q$, and then supposing  $\ze=\ze_ke^{ikq}$ to eliminate $q$ (which is $\theta$ in the previous language) in lieu of the wavenumber $k$,   $\ze_k(p,t)$ becomes our dynamical variable that satisfies 
\[
\ze_{k,t} = - i k p \ze_k + i k f_0'  \phi_k,
\quad\mathrm{and}\quad
 \phi_k  = {k^{-2}}\int_{\R} \!\! \dint \bar{p} \,  \ze_k(\bar{p},t)
\,,
\]
which simplifies to the following integro-differential equation for $\ze_k$:
\bq
\ze_{k,t}=- i kp\ze_k+ {if_0'} \,k^{-1}\!  \int_{\mathbb{R}}\!d\bar{p}\, \ze_k(\bar{p},t)=:-\T_k \ze_k\,.
\label{zek}
\eq
Here we have introduced the time evolution operator $\T_k$, whose spectrum under changes in $f_0$ we will study to understand the CHH bifurcation.

The linearized equations inherit  a Hamiltonian structure.  Because of the noncanonical form, linearization requires expansion of the Poisson bracket as well as the Hamiltonian. In terms of the variables $\ze_k$ and $\ze_{-k}$  the Hamiltonian is:
\[
H_L[\ze_k,\ze_{-k}] = \frac{1}{2}\sum_{k\in \N}\left(-\int_{\mathbb{R}}\!dp\, \frac{p}{f_0'}\, |\ze_k|^2+
|\phi_k|^2\right)\,,
\]
with the Poisson bracket
\begin{align}
\{F,G\}_L &= \sum_{k\in \N}ik\int_{\mathbb{R}}\!dv\, f_0'\left(\frac
{\delta F}{\delta \ze_k}\frac{\delta G}{\delta \ze_{-k}}-\frac{\delta F}{\delta \ze_{-k}}
\frac{\delta G}{\delta \ze_k}\right)\,.
\label{pbk}
\end{align}
Observe from (\ref{pbk}) that $k\in\N$ and thus  $\ze_k$ and $\ze_{-k}$ are independent variables that are almost canonically conjugate. Thus the complete system is
\[
\ze_{k,t}=-\T_k \ze_k\qquad{\rm and}\qquad \ze_{-k,t}=-\T_{-k} \ze_{-k}\,,
\]
whence it can be shown directly  that the spectrum is Hamiltonian.

Now we consider properties of the evolution operator $\T$  defined by (\ref{zek}).  
We suppose $\ze_k$ varies as $\exp(-i\om t)$, 
where $\om$ is the frequency and $i\om$ is the eigenvalue.  For convenience we also use $u:=\om/k$, where 
we can view $k\in \R^{>0}$ by varying the system size.   The system is said to be spectrally stable if the  spectrum of $\T$ is less than 
or equal to zero or the frequency is always in the closed lower half plane. Since the system  is Hamiltonian,  
the question of stability reduces to deciding if the spectrum is confined to the imaginary axis.   
The solutions of a spectrally stable system are guaranteed to grow at most sub-exponentially.

The operator $\T_k$ is the sum of a multiplication  operator and an integral operator, and the multiplication 
operator causes the continuous spectrum to be composed of the entire imaginary axis except possibly for some discrete points.  
Instability comes from  the point spectrum. The linearized Vlasov Poisson equation is not spectrally stable when the time evolution operator has an element of the point spectrum away from the imaginary axis (implying a doublet or quartet of modes with non-trivial real part). 
The point spectrum are the roots of the plasma dispersion function:
\[
\vep(k,u):=1-\frac1{k^{2}}\int_{\R}\!dp 
\frac{f_0'}{p-u}\,.
\]
Here $u=\omega/k$.
The one-dimensional linearized Vlasov-Poisson system with homogeneous equilibrium $f_0$ is spectrally unstable 
if for some $k\in\R^{>0}$ and $u$ in the upper half plane,  the plasma dispersion function vanishes.

Using the Nyquist method  that relies on the argument principle of complex analysis,  Penrose \cite{penrose} was able to relate the vanishing of $\vep(k,u)$ to the winding number  of the closed curve determined by the real and imaginary parts of $\vep$ as $u$ runs along the real axis.  Such closed curves are called Penrose plots.  The crucial quantity is the  integral part of $\vep$ as $u$ approaches the real axis from above:  
\[
\lim_{u\rightarrow 0^+}\frac{1}{\pi}\int_{\R}\!dp\, \frac{f_0'}{p-u}=\mathcal{H}[f_0'](u)+if_0'(u)\,,
\]
 where $\mathcal{H}[f_0']$ denotes the Hilbert transform, $\mathcal{H}[f_0']=\frac{1}{\pi}\dashint\!dp\,{f_0'}/({p-u})$, where $\dashint:=PV\!\!\int_{\R}$   indicates the Cauchy principle value, leading to the following expression for
the contour, parametrized by $u\in\R$, in the complex plane:
\[
\vep(k,u):=1-\pi k^{-2}  \mathcal{H}[f_0'](u)-i \pi k^{-2} f_0'(u)\,.
\]
The image of the real line under this mapping is  the Penrose plot, and its winding number about the origin is the number of members of the point spectrum of $T_k$ in the upper half plane.

\begin{figure}[htbp]
\begin{center}
\includegraphics[scale=.61]{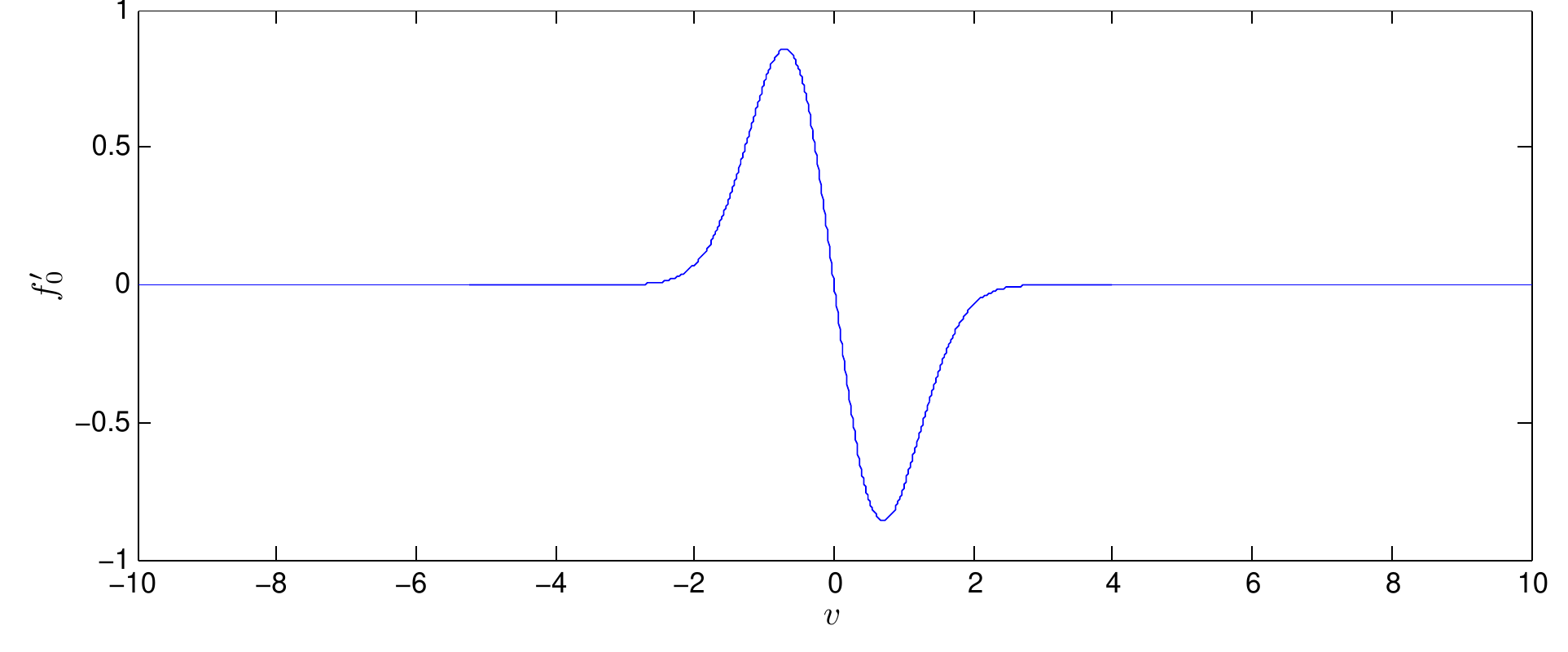}
\end{center}
\caption{$f_0'$ for a Maxwellian distribution function.}
\label{gaus1}
\end{figure}

\begin{figure}[htbp]
\begin{center}
\includegraphics[scale=.55]{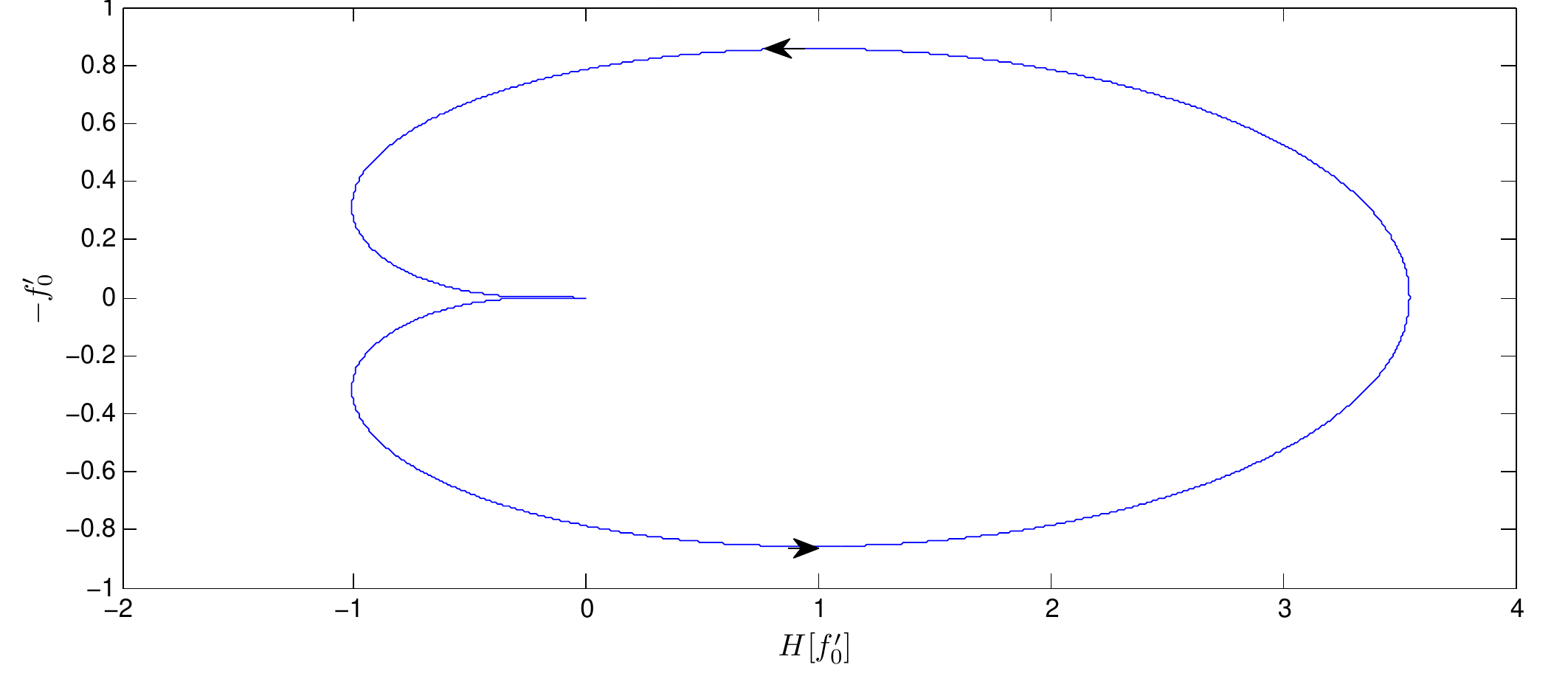}
\end{center}
\caption{Stable Penrose plot for a Maxwellian distribution function.}
\label{penroseGaus}
\end{figure}

Figure \ref{gaus1}  shows the derivative of the distribution function, $f_0'$,   for the case of a Maxwellian distribution $f_0=e^{-p^2}$ and Fig.~\ref{penroseGaus}  shows  the contour $-\mathcal{H}[f_0']-if_0'(u)$ that emerges from the origin in the complex plane at $u=-\infty$,  descends, and then wraps  around to return to the origin at $u=\infty$.  From this figure it is evident that the winding number of the $\vep(k,u)$-plot  is zero for any fixed $k\in\R$, and as a result there are no unstable modes. Here we take the value of $k$ to be fixed.

\bigskip

\begin{figure}[htb]
  \begin{center}
    \subfigure[]{\includegraphics[scale=.55]{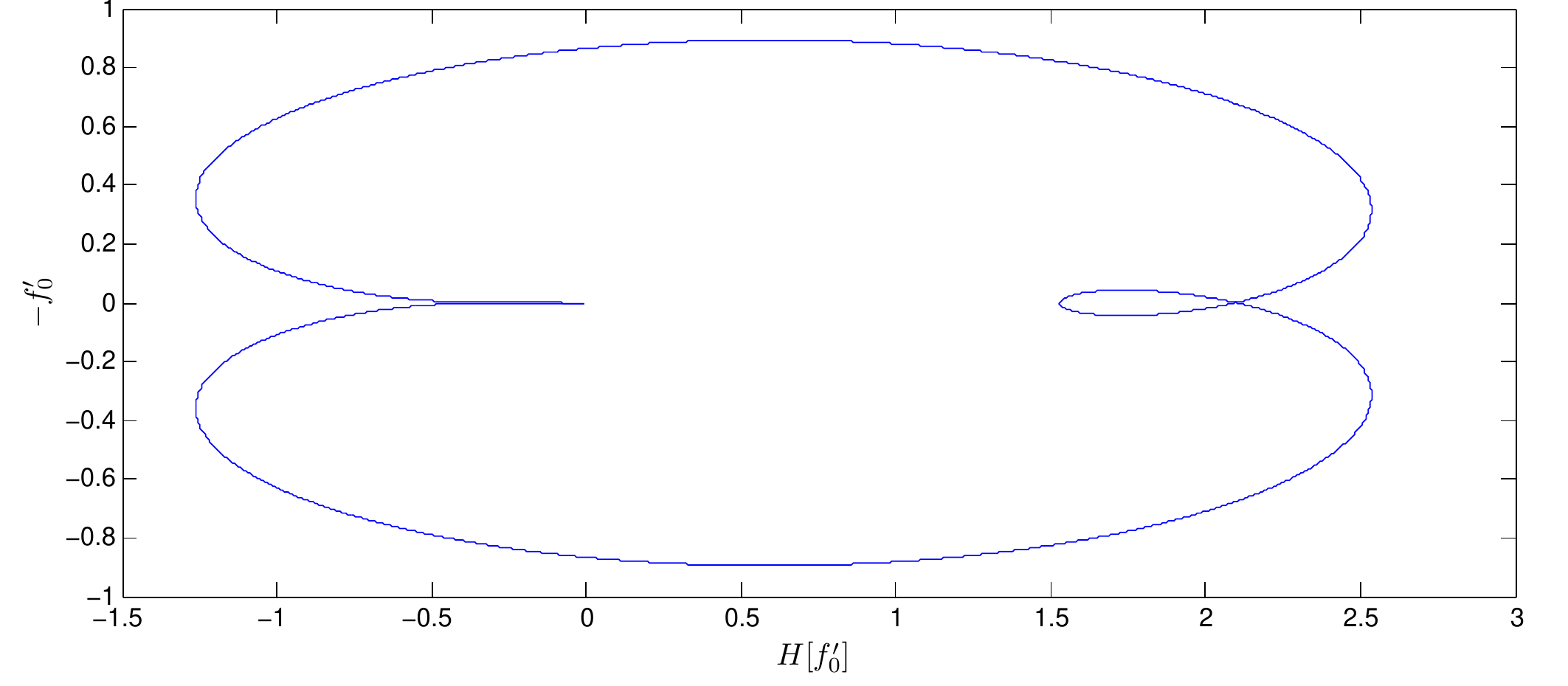}}
    \subfigure[]{\includegraphics[scale=.55]{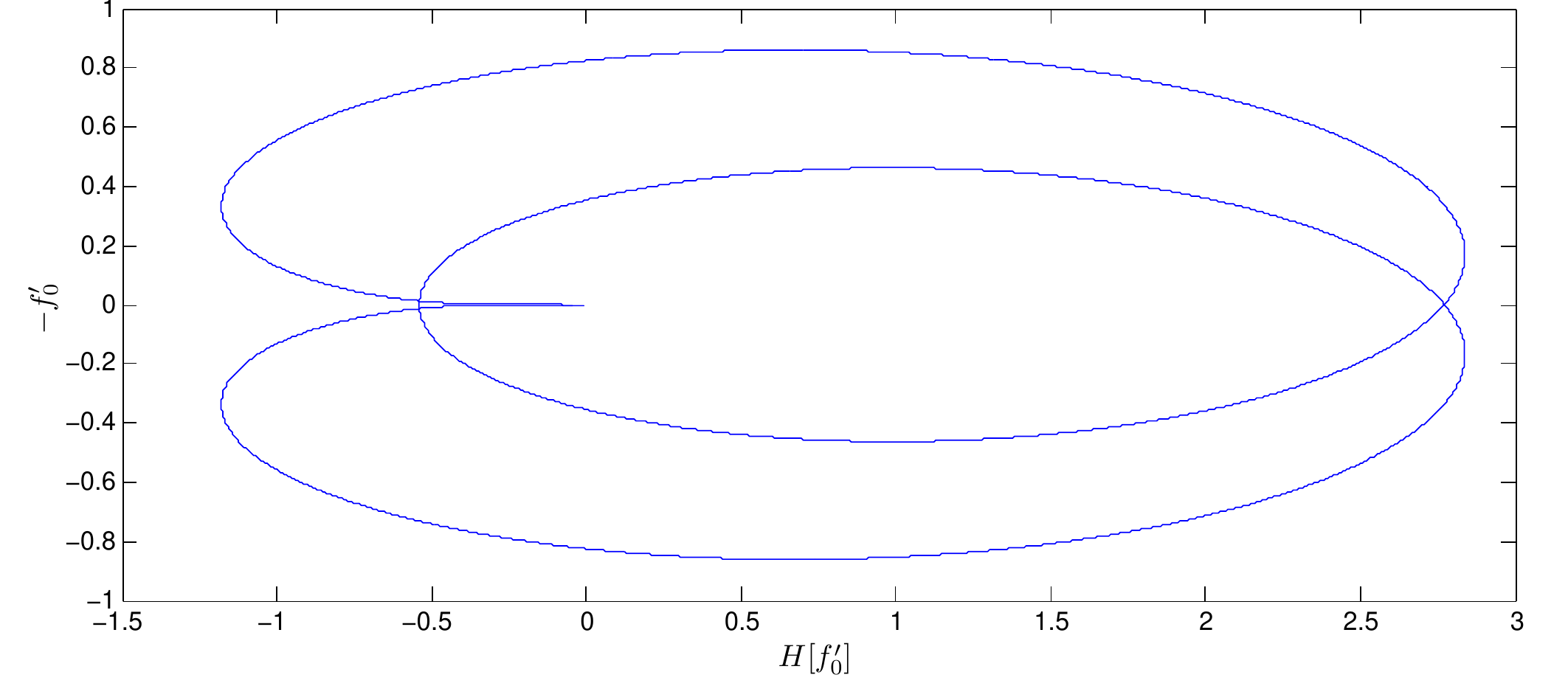}}
  \end{center}
  \caption{Penrose plot for a superposition of Maxwellian distribution functions with a) a  stable separation 
  and  b)  an unstable separation.}
\label{2max}
\end{figure}

Penrose plots can be used to visually determine   spectral stability.  As described above, the  Maxwellian distribution function is stable, as the resulting $\varep$-plot does not encircle the origin. However, it is not difficult to 
construct unstable distribution functions. In particular,   the superposition of two displaced Maxwellian distributions, 
$f_0=e^{-(p+c)^2}+e^{-(p-c)^2}$, is such a case. 
As $c$ increases the distribution goes from stable to unstable. This instability is known as the two-stream instability.  Figures~\ref{2max}a and  \ref{2max}b demonstrate how the transition from stability to  instability is manifested in a Penrose plot. 

\bigskip

At the bifurcation point the Penrose plot crosses the origin, indicating the vanishing of the dispersion relation on the real axis and therefore the presence of a member of the point spectrum. This eigenmode will be stable because $u\in \R$, and will be embedded within the continuous spectrum. Thus,  the two-stream instability is an example of the CHH bifurcation. 

The description of the CHH  bifurcation requires that one be able to assign an energy signature to the continuous spectrum.  Because eigenfunctions associated with continuous spectra are not normalizable, this requires some delicacy.  This was first done in the Vlasov context in  \cite{MP92}, where a comparison to the usual energy signature for discrete modes was given, and  followed by a rigorous treatment of signature in \cite{morrison00a}.  In the  context of shear flow,  signature was defined  in \cite{BM02}, in magnetofluids in \cite{hirota},   and for the general system described in the present subsection in \cite{morrison03}.   A rigorous version of  Kre\u{i}n's theorm for the CHH  bifurcation was given in \cite{hagstrom}. We shall give a general description of this energy signature for the continuous spectrum in \cite{chaptII},  but we motivate it  here first by treating the analogous version of this instability in the context of the waterbag model, which will have the advantage of only possessing a discrete spectrum.

\subsubsection{Bifurcations in the waterbag model: Vlasov interpretation}
\label{sssec:WB}

One important feature of the system  (\ref{eq:den.eq}) is that its solution is a symplectic rearrangment of the initial condition 
$\mathring{f}(q,p)=f(q,p,0)$, i.e.,  its solution has the form  
\begin{equation}
f(q,p,t)=\mathring{f}\circ \mathring{z}(q,p,t)\,, 
\end{equation} 
where $\mathring{z}(q,p,t)=(\mathring{q}(q,p,t),\mathring{p}(q,p,t))$ is a canonical transformation.

The rearrangement comes from the solution of the ordinary differential equation for a single particle in the self-consistent potential $\phi$.  This implies that the level set topology of the initial condition is preserved, which can be leveraged to simplify the equations in the case of certain types of initial conditions.   One such simplification is known as the waterbag reduction (see, e.g., \cite{berk}), in which it is assumed that the initial condition $\mathring{f}$ is a sum of characteristic functions. This property is preserved under composition with the symplectic map $\mathring{z}$, so that the solution remains a sum of characteristic functions. The equations simplify to equations for the locations of the contours separating different regions of constant $f$. 
Piecewise constant initial conditions lead to a fluid closure that is exact for waterbag initial conditions, and the $1+1$ theories in the previous section can be seen to arise  from such an ansatz. We will exploit the reduction by using a layered waterbag or onion-like initial condition to closely  approximate a continuous distribution function that undergoes the bifurcation to linear instability we are interested in.  In this way we will be able to connect the HH bifurcation with the CHH  bifurcation which we describe later. 

We begin by assuming  $f$ to be  piecewise constant between $M$ curves $p_{\al}(q,t)$, i.e., 
\[
f(q,p,t)= f_{\al} \qquad{\rm if}\qquad p_{\al}<p<p_{{\al}+1}
\]
where $f_{\al}$ is a positive constant.   The equations for the curves $p_{\al}$ come from the equations of single particle motion for a particle at $(p_{\al}(q,t),q)$:
\[
p_{{\al},t}+p_{\al}\, p_{{\al},q}=-\phi_q 
\quad\mathrm{and}\quad
\phi_{qq}=1-\sum_{\al} f_{\al}(p_{{\al}+1}-p_{\al})\,, 
\]
and this system is Hamiltonian, with  Hamiltonian function being the classical energy:
\[
H= \sum_{\al}  \frac{\Delta f_{\al}}{6}  \int_{\mathbb{T}} dq \, p_{\al}^3+ \frac{1}{2}\int_{\mathbb{T}} dx\left(\phi_q\right)^2\,.
\]
Here $\Delta f_{\al}=f_{{\al}-1}-f_{\al}$, is the change in the distribution function when crossing the ${\al}$th waterbag layer.
The Poisson bracket is similar to those seen in Hamiltonian fluid theories \cite{morrison98}:
\[
\{F,G\}=\sum_{\al}\int_{\mathbb{T}} \frac{dq}{\Delta f_{\al}} \, \frac{\delta F}{\delta p_{\al}}\p\frac{\delta G}{\delta p_{\al}}
\]

The equilibria of the waterbag model that we are interested in studying are charge neutral and spatially homogeneous, $p_{\al}=p_{{\al}0}$ constant, such that  the  electric potential $\phi\equiv 0$. We chose such a state and linearize about it, yielding the  equations of motion 
\[
p_{{\al},t}+p_{{\al}0}\, p_{{\al},q}=-\phi_q 
\quad \mathrm{and}\quad
\phi_{qq}=-\sum_{{\al}} f_{\al}(p_{{\al}+1}-p_{\al})\,.
\]
Moving  to Fourier space and  eliminating the dependence on $q$ in favor of the wavenumber $k$ gives
\[
p_{k,t}^{\al}+ik\, p_0^{\al}  p_k^{\al}={i}{k}\sum_{\al} f_{\al}(p_k^{{\al}+1}-p_k^{\al})\,, 
\]
the equations of motion for the Fourier coefficients.   In terms of the Fourier coefficients, the Hamiltonian of the linearized system is 
\[
H=-\frac1{2}\sum_{k\in\Z}\left(\sum_{\al} p_0^{\al} {\Delta f_{\al}}\, |p_k^{\al}|^2 
+   {k^2} |\phi_k|^2
\right)\,.
\]
Here the term $-p_0^{\al} \Delta f_{\al}$ arises from  the
term $-pf_0'$ in the linearized Vlasov equation, which indicates the signature of the continuous spectrum.
The bracket is the bracket of the original nonlinear system written in terms of the Fourier modes:
\[
[f,g]=\sum_{k\in \N}\sum_{\al}\frac{ik}{\Delta f_{\al}}\, 
\left( \frac{\p f}{\p p_k^{\al}}\frac{\p g}{\p p_{-k}^{\al}}
- \frac{\p g}{\p p_k^{\al}}\frac{\p f}{\p p_{-k}^{\al}}
\right)\,.
\]
This bracket is nondegenerate and therefore the system is nearly canonical in terms of the new variables.
In particular, for a given pair $k,-k$, the linear equations form a finite-dimensional canonical Hamiltonian system with scaling similar to that of Sec.~\ref{ssec:mfexamples}.

 The dispersion relation for this system, for a given wave number $k$, and $u=\omega/k$,
is derived by multiplying the $i$th equation by $\Delta f_{\al}$ and summing, which is analogous to that for the Vlasov system, 
\[
\varepsilon(u,k)=1-\frac{1}{k^2}\sum_{\al} \frac{\Delta f_{\al}}{u- p_0^{\al}}=0\,.
\]
This dispersion relation can be analyzed graphically in terms of $u$. There are poles of the dispersion function
 where $u=p_0^{\al}$. For $u\in(p_0^{\al},p_0^{{\al}-1})$, the
dispersion function always has a zero if $\Delta f_{{\al}+1}$ has the same sign as $\Delta f_{\al}$, 
because $\epsilon$ will converge to the opposite value of infinity at each end of the interval. Therefore there will be at least one zero in each interval that has this property. In intervals where $\Delta f_{{\al}+1}$ and $\Delta f_{\al}$ have different signs there are either $0$ zeros or an even number of zeros, because $\varepsilon$ must converge to the same value of infinity.


The reader may have noticed a similarity between the above formulas, and the multi-fluid formulas of 
Sec.~\ref{ssec:mulitftheo}.  In fact, the waterbag models are  examples of   multi-fluid models, which are thus 
exact fluid closures of the Vlasov-Poisson system. This can be seen by writing the
waterbag model in terms of  new variables $\rho_{\al}$ and $u_{\al}$ given by 
\[
\rho_{\alpha}=\ \left(p_{\alpha}-p_{\alpha-1}\right)/ f_{\alpha}\quad \mathrm{and}\quad
u_{\alpha}= \left(p_{\alpha}+p_{\alpha-1}\right)/2\,, 
\]
where $\rho_{\alpha}$ is a fluid density, and $u_{\alpha}$ is a fluid velocity. Under this change of variables 
the equations governing the waterbag model take the following form:
\bqy
&&\rho_{\alpha,t}+ \left(u_{\alpha} \rho_{\alpha}\right)_{q} =0\,, \quad \phi_{qq}=-\sum_{\al} \rho_{\alpha}
\nonumber\\
 &&\left(\rho_{\alpha} u_{\alpha}\right)_{t}+ 
 \left(u_{\alpha}^2\rho_{\alpha}+ {\rho_{\alpha}^3f_{\alpha}^2}/{12}\right)_{q}=-\rho_{\alpha}\phi_q \,.
 \nonumber
\eqy
Evidently, under the identification $p_{\alpha}={\rho_{\alpha}^3 f_{\alpha}^2}/{12}$ 
(or $h_{\alpha}=\rho_{\alpha}^2 f_{\alpha}^2/{8})$, the above equations are identified as a multi-fluid Hamiltonian system. 

The dispersion function can also be rewritten in terms of the new variables, so that it resembles that analogous expressions (i.e., the diagravic or dielectric functions of the multi-fluid section).  After linearizing  around an equilibrium state with $\rho_0^{\alpha}$, $u_0^{\alpha}$, and then performing some algebraic manipulations yields 
\bq
\varepsilon(k,\omega)= 
1  -\sum_{\alpha} 
\frac{\rho_0^{\alpha}}{\left(
\omega- k u_0^{\alpha}\right)^2 
- k^2(u^{\al}_{\th})^2}\,,
\label{wbvareps}
\eq
where $u^{\al}_{\th}:=\sqrt{{(\rho_0^{\alpha})^2}/{f_{\alpha}^2}}$ is a thermal velocity that measures the width in velocity  space of a waterbag.  Thus bifurcations in the waterbag model, the Vlasov-Poisson system, and Hamiltonian multi-fluid equations are all described using similar mathematical expressions.


Because the waterbag system is a finite-dimensional canonical linear Hamiltonian system, the standard results of that theory apply, including Kre\u{i}n's theorem. We can therefore determine whether there are any unstable
modes by counting the total number of neutral eigenvalues. If it is equal to the number of degrees of freedom of the system, then we can expect stability, otherwise, due to the fact that eigenvalues off the
imaginary axis come in quartets, we can expect instability. 

Now we determine the signature of each of the stable discrete modes of the waterbag model. Beginning with the linearized equations, and assuming the normalization condition, $
1= \sum_{\al} \Delta f_{\al} \,  p_{\al}/k=-k\phi_k$, 
we find  the Fourier eigenvector  $p_{k}^{\al}={1}/[{k}(p_{0}^{\al} -u )]$.  Using this in the expression for the energy, we get a formula for the energy of a discrete mode, viz.
\bq
H=-\sum_{\al} \frac{p_{0}^{\al}}{2k^2}\frac{\Delta f_{\al}}{(p_{0}^{\al}-u)^2}+\frac{1}{2}\,.
\label{Hwb}
\eq
Next,  replacing $p_{0}^{\al}$ in the numerator of (\ref{Hwb}) by  $p_{0}^{\al}=u+(p_{0}^{\al}-u)$ leads to
\bq
H=\frac{1}{k^2}\sum_{\al}\left(\frac{\Delta f_{\al}\,  p_{0}^{\al}}{(u-p_{0}^{\al})^2}+\frac{\Delta f_{\al}}{p_{0}^{\al}-u}\right)+\frac{1}{2} 
= \frac{1}{k^2}\sum_{\al}\frac{\Delta f_{\al}\, p_{0}^{\al}}{(u-p_{0}^{\al})^2} = u\frac{\partial \varepsilon}{\partial u}\,,
\nonumber
\eq
where in the last expression we obtain the dielectric energy (with the electric field amplitude dependence scaled away).

\begin{figure}[htb]
  \begin{center}
    \subfigure[]{\includegraphics[scale=.55]{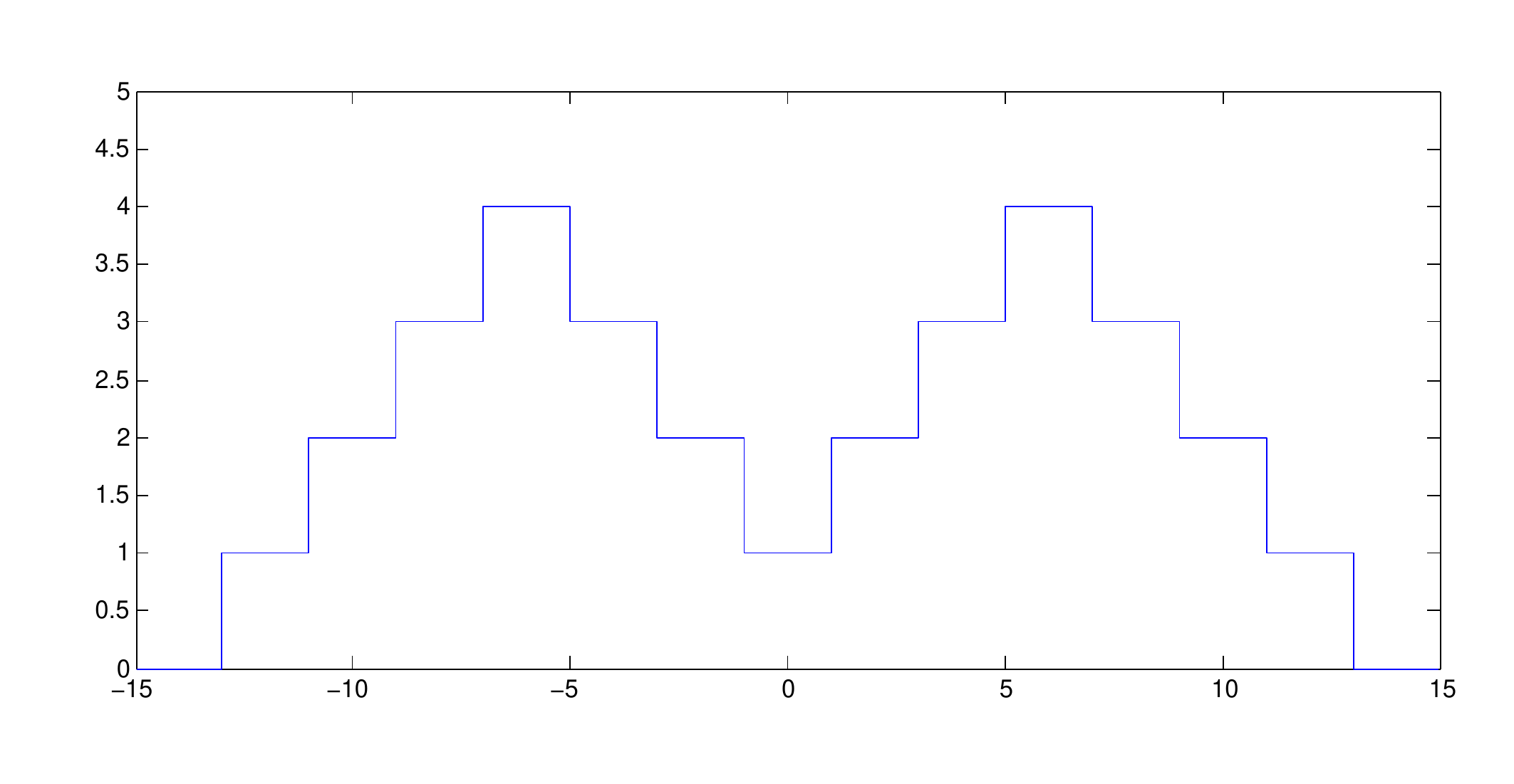}}
    \subfigure[]{\includegraphics[scale=.55]{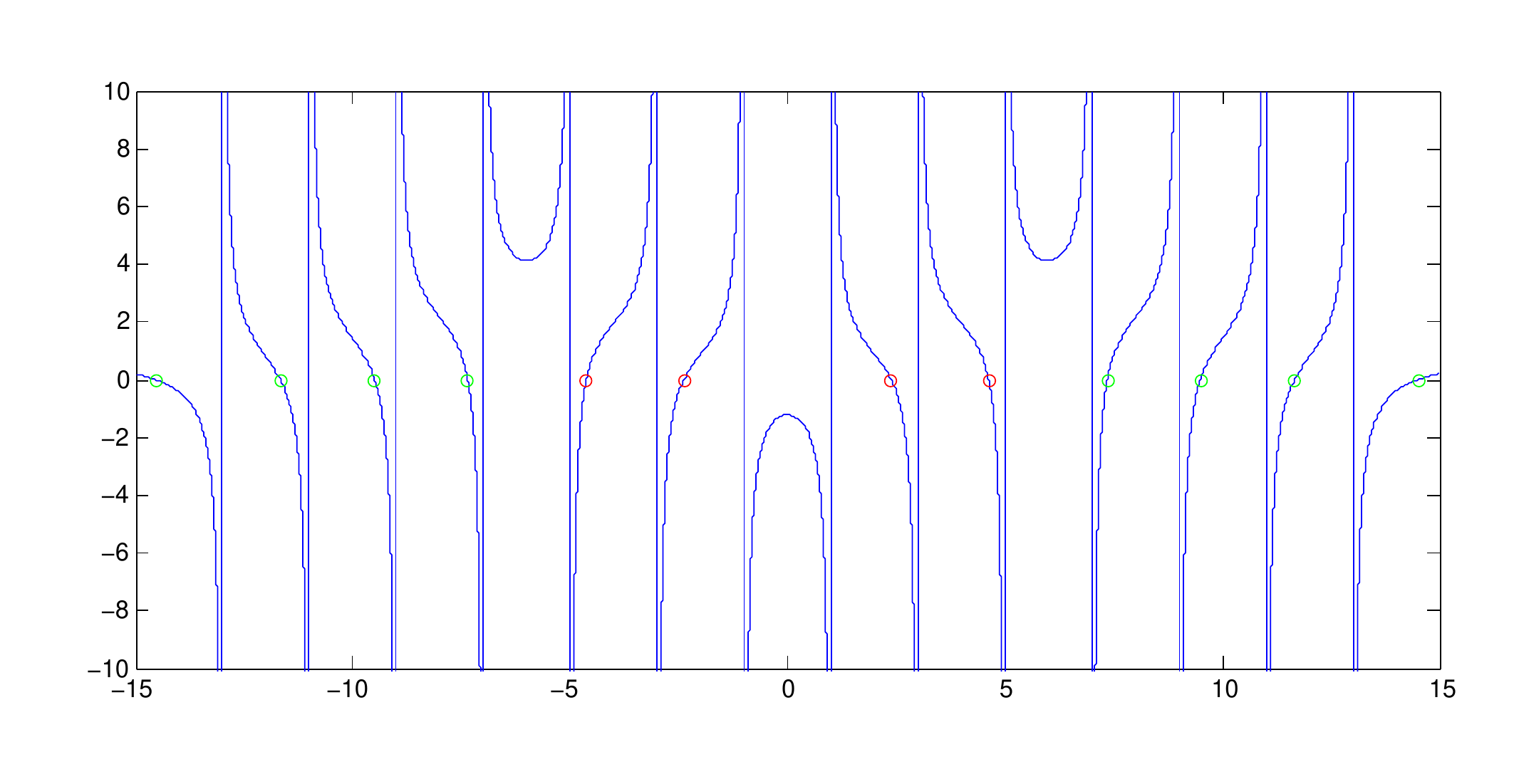}}
  \end{center}
  \caption{a) Multi-level waterbag distribution.  b) Plot of corresponding  dispersion relation, with positive and negative signature modes indicated by green and red circles, respectively.}
 \label{waterbag}
\end{figure}

The energy of a discrete mode is proportional to  the derivative of the dispersion function at the frequency corresponding to
the mode. As mentioned previously, this  familiar formula is also true for embedded modes in the Vlasov equation  \cite{sm94}, and is particularly convenient for use in the waterbag model because it allows geometric evaluation of the signature of modes in the waterbag model.  Suppose at first that $u>0$. Then the signature of a mode is positive if the dispersion function is increasing at the mode, and negative if it is decreasing at the mode. If $\Delta f_{\al}$ does not change sign from one interval to the next, and there is one mode in the corresponding interval, the mode will have signature $-\sgn(p_{0}^{\al}\Delta f_{\al})$. Similarly, any modes in the same interval must have opposite signature (or one must have zero signature), because the dispersion function must cross the axis in opposite directions at each discrete mode. An example of such a waterbag distribution function is plotted in Fig.~\ref{waterbag}a, and the dispersion relation is plotted in Fig.~\ref{waterbag}b, where we have marked the zeros  with colored circles that indicate their signature. As noted, in the general case there is exactly $1$ mode in intervals where  $\Delta f_{\al}$ does not change sign, and either $0$ or $2$ modes in intervals where it does change sign.

\begin{figure}[htbp]
\begin{center}
\includegraphics[scale=.55]{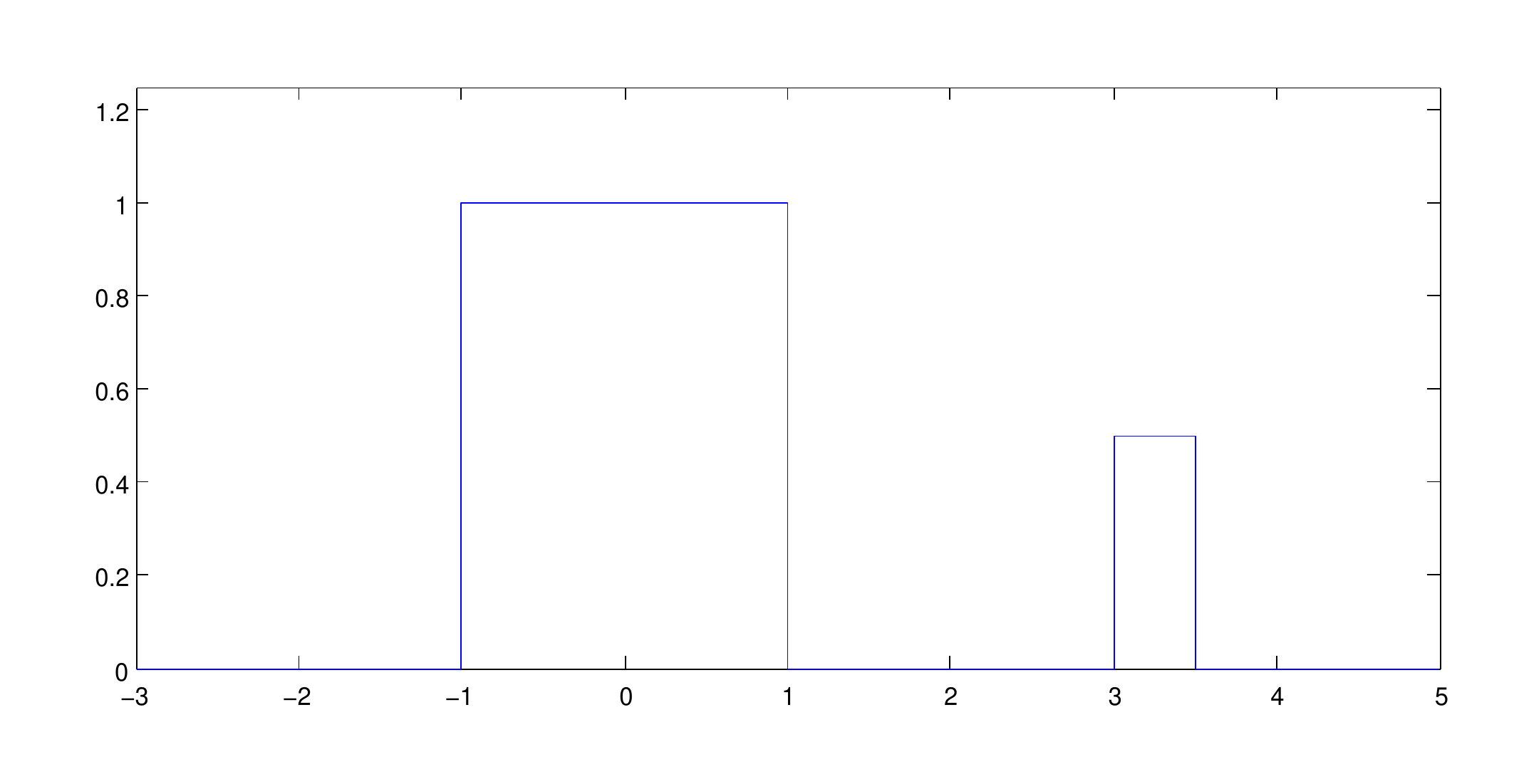}
\end{center}
\caption{Plot of a waterbag distribution function meant to capture the electron two-stream instability. As the small waterbag is moved closer to the large one,  a positive energy mode will collide with a negative energy mode and give rise to the two-stream or bump on tail instability.}
\label{twostreamf0}
\end{figure}

\begin{figure}[htpb]
  \begin{center}
    \subfigure[]{\includegraphics[scale=.542]{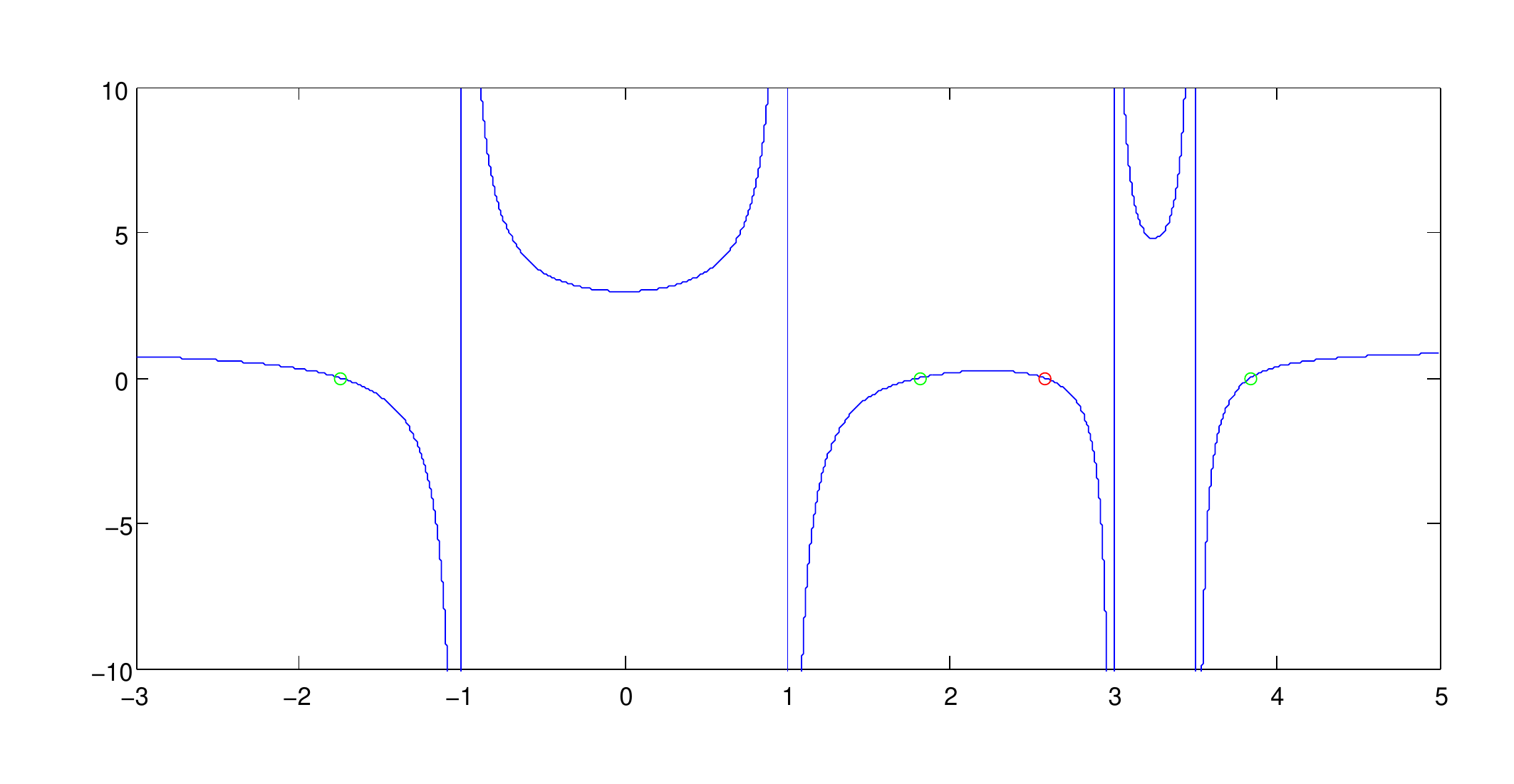}}
    \subfigure[]{\includegraphics[scale=.542]{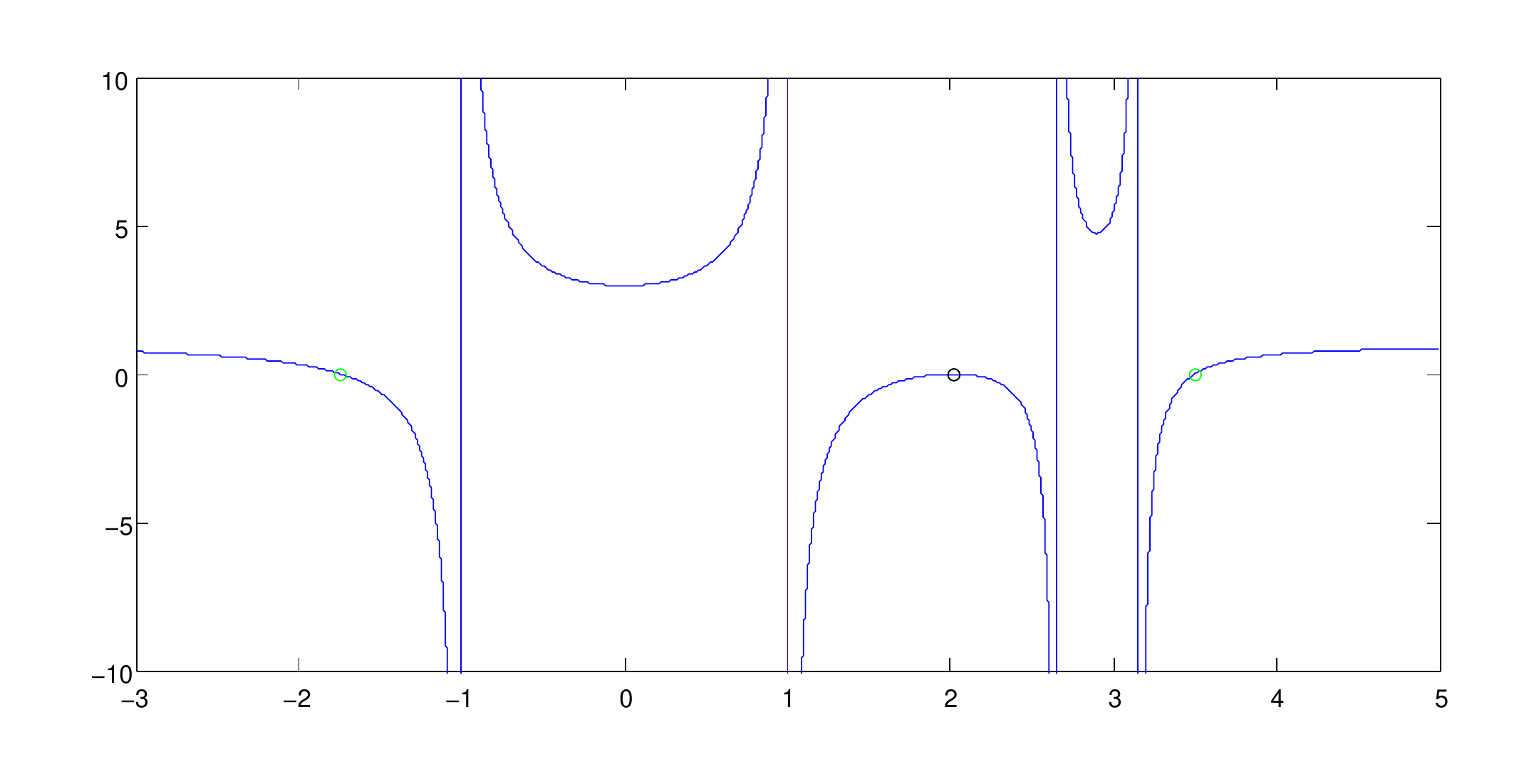}}
    \subfigure[]{\includegraphics[scale=.542]{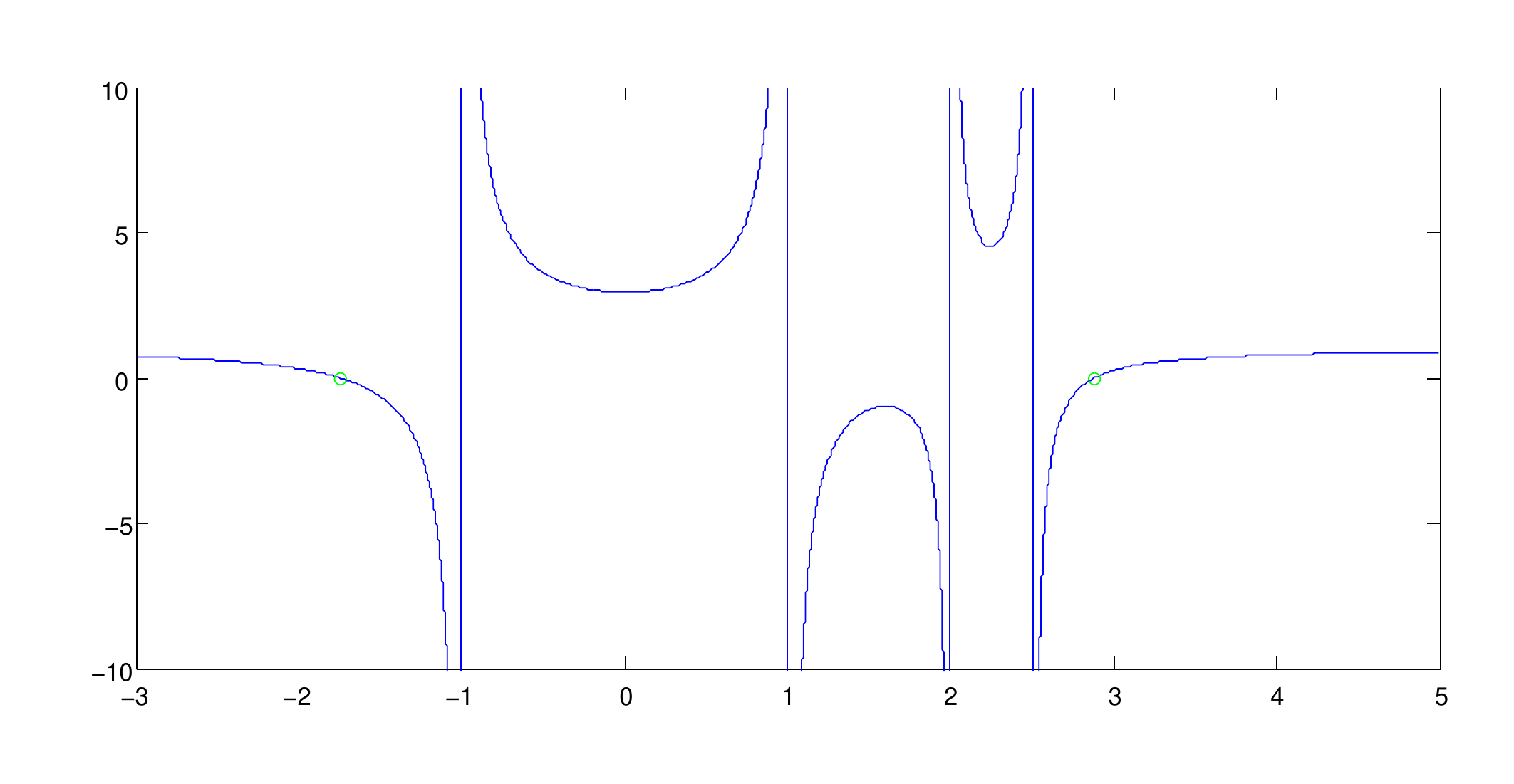}}
  \end{center}
  \caption{Dispersion function for a  two-stream distribution function for parameter values corresponding to  a) stable, b) neutral, and c) unstable equilibria.}
\label{2stream}
\end{figure}

Using the waterbag model we can replicate the most important instabilities of the Vlasov-Poisson equation,
in particular the two stream instability and bump on tail instability. Both of these instabilities can be
emulated by a waterbag model with only a few 'layers' (fluids). In particular, we will consider the special case of a waterbag with 5 layers as depicted in Fig.~\ref{twostreamf0}.  Observe, the outermost two have vanishing distribution function,  i.e.,  $f_1=f_5=0$, while we choose $f_2=1$, $f_3=0$, and $f_4=.5$, so that the  distribution has  two peaks, one large and one small, separated by a valley.  The stability of this model depends on the various parameters involved in defining the equilibrium. For a very large separation of the two peaks, the two-stream distribution function will be stable as depicted in Fig.~\ref{2stream}a; as the peaks are moved closer together the two modes in the valley of the distribution function between the two peaks move closer together, eventually colliding as depicted in 
Fig.~\ref{2stream}b, and leaving the axis to become a pair of exponentially growing and decaying modes as depicted in Fig.~\ref{2stream}c.

 This transition here is identical to that which occurs in the two stream instability of the Vlasov equation (or the corresponding bump on tail instability). In the waterbag case there is a positive energy mode that collides with a negative energy mode in the  valley of the distribution function.

\section{Summary and conclusions}
\label{sec:conclu1}

In this chapter we have described bifurcations in general classes of noncanonical Hamiltonians systems that describe, e.g.,  matter as fluid or kinetic theories.  In the multi-fluid systems of Sec.~\ref{sec:Discrete} we showed how to linearize, canonize and, for stable systems with discrete spectra, diagonalize to obtain a normal form.  Hamiltonian bifurcations to instability were  described,  examples of SS bifurcations were given, but the emphasis was on the HH bifurcation.  From the normal form, signature was  identified, and it was seen that Kre\u{i}n's theorem applies, just as for finite-dimensional systems.  Next,  the class of 2+1 Hamiltonian theories of Sec.~\ref{sec:theories} were defined and considered.  These theories generically posses continuous spectra when linearized,  but the specific case of the Vlasov-Poisson systems was treated in detail.  In particular,  Penrose plots, which allow one to describe transitions to instability, via embedded modes in a continuous spectrum, were described.  The technique here is  of general utility, e.g.,  it was worked out also in detail for shear flow in \cite{BM98}.  It was also shown how to canonize the linearization of these 2+1 theories.  Next, in order to understand the relationship between discrete  bifurcations and the CSS and CHH bifurcations, we introduced the waterbag model, which  is a  reduction of the 2+1 class to  a class of systems with a  countable number of degrees of freedom, in which the continuous spectrum is discretized.  The identification of the waterbag models with the multi-fluid models of  Sec.~\ref{sec:Discrete} was made  and, consequently, the procedure for  canonization and diagonalization of the waterbag models was established.   

A main motivation for studying Hamiltonian systems is their universality, i.e., one is interested in understanding features of one system that apply to all. In this chapter we have shown how infinite-dimensional noncanonical Hamiltonian systems enlarge this universality class.  It is clear that the same bifurcations occur in a variety of systems that describe different physical situations. Any specific system within our classes of systems may possess SS bifurcations, positive and negative energy modes, and Kre\u{i}n's theorem for HH bifurcations.  Our aim is  show that an analogous situation transpires  for CSS and CHH bifurcations.  However, continuous spectra  are  harder to deal with mathematically and  functional analysis is essential, but the existence of analogous behavior in the  cases considered here guides us to develop a theory.   For example, one can interpret the CHH bifurcation as an  HH bifurcation with the  second mode coming from the continuous spectrum.    As stated before, the contents of this chapter are to  set the stage for the explicit treatment of bifurcations with the continuous spectrum of  \cite{chaptII}, to which we direct the reader.

\bigskip
\noindent \textbf{Acknowledgements.}
 Hospitality of the GFD Summer Program, held at the Woods Hole Oceanographic Institution, is greatly appreciated.  PJM and GIH were supported by USDOE grant nos.~DE-FG02-04ER54742 and DE-FG02-ER53223, respectively.


%

\end{document}